\documentclass[twocolumn,aps,prl,reprint,
superscriptaddress
]{revtex4-1}
\usepackage[pdftex]{graphicx}
\usepackage{amsmath,amssymb}
\usepackage[pdftex,
            colorlinks=true,
            citecolor=blue,
            linkcolor=blue]{hyperref}

\graphicspath{{./fig/}{./fig_phase_diagram_t20/}{./fig_phase_diagram_t202/}}

\usepackage{braket}
\usepackage{times,multirow,amsfonts,bm,xspace,pifont}
\usepackage{soul}
\usepackage[normalem]{ulem}

\begin{document}
\newcommand{\rr}{{\bm r}}
\newcommand{\q}{{\bm q}}
\renewcommand{\k}{{\bm k}}
\newcommand*\wien    {\textsc{wien}2k\xspace}
\newcommand*\textred[1]{\textcolor{red}{#1}}
\newcommand*\textblue[1]{\textcolor{blue}{#1}}
\newcommand{\ki}[1]{{\color{red}\st{#1}}}
\newcommand{\sgn}{\mathrm{sgn}\,}
\newcommand{\tr}{\mathrm{tr}\,}
\newcommand{\Tr}{\mathrm{Tr}\,}
\newcommand{\GL}{{\mathrm{GL}}}
\newcommand{\talpha}{{\tilde{\alpha}}}
\newcommand{\tbeta}{{\tilde{\beta}}}
\newcommand{\mathN}{{\mathcal{N}}}
\newcommand{\mathQ}{{\mathcal{Q}}}
\newcommand{\bv}{{\bar{v}}}
\newcommand{\bj}{{\bar{j}}}

\newcommand{\YY}[1]{\textcolor{magenta}{#1}}
\newcommand{\KN}[1]{\textcolor{green}{#1}}
\newcommand*{\KNS}[1]{\textcolor{green}{\sout{#1}}}
\newcommand*\YYS[1]{\textcolor{magenta}{\sout{#1}}}
\newcommand{\reply}[1]{\textcolor{red}{#1}}
\newcommand{\replyS}[1]{\textcolor{red}{\sout{#1}}}

\title{Magnetism and superconductivity in mixed-dimensional periodic Anderson model for UTe$_{2}$}

\author{Ryuji Hakuno}
\email[]{hakuno.ryuji.46v@st.kyoto-u.ac.jp}
\affiliation{Department of Physics, Graduate School of Science, Kyoto University, Kyoto 606-8502, Japan}
\author{Kosuke Nogaki}
\affiliation{Department of Physics, Graduate School of Science, Kyoto University, Kyoto 606-8502, Japan}
\author{Youichi Yanase}
\affiliation{Department of Physics, Graduate School of Science, Kyoto University, Kyoto 606-8502, Japan}
\date{\today}

\begin{abstract}
UTe$_{2}$ is a strong candidate for a topological spin-triplet superconductor, and it is considered that the interplay of magnetic fluctuation and superconductivity is essential for the origin of the superconductivity. 
Despite various experiments suggesting ferromagnetic criticality, neutron scattering measurements observed only antiferromagnetic fluctuation and called for theories of spin-triplet superconductivity near the antiferromagnetic quantum critical point.
We construct a periodic Anderson model with one-dimensional conduction electrons and two- or three-dimensional $f$-electrons, reminiscent of the band structure of UTe$_2$, and show that 
ferromagnetic and antiferromagnetic fluctuations are reproduced depending on the Fermi surface of $f$ electrons. These magnetic fluctuations cooperatively stabilize spin-triplet $p$-wave superconductivity. 
We also study hybridization dependence as a possible origin of pressure-induced superconducting phases and find that moderately large hybridization drastically changes the antiferromagnetic wave vector and stabilizes $d$-wave superconductivity.
\end{abstract}
\maketitle

\textit{Introduction}. --- 
Spin-triplet superconductivity is an exotic quantum condensed state of matter.  Interest in spin-triplet superconductors is growing as they are platforms of multi-component superconductivity, bulk topological superconductivity~\cite{Schnyder2008,Sato2010,Fu-Berg2010,sato-fujimotoreview}, and superconducting spintronics~\cite{Linder2015}.
Since the discovery of superconductivity~\cite{Ran2019}, UTe$_2$ has been considered to be a spin-triplet superconductor, and intensive studies have been devoted to clarifying the superconducting states in UTe$_2$~\cite{Aoki2022review}. 
One of the authors previously proposed the existence of topological superconductivity in UTe$_2$~\cite{Ishizuka2019}.
However, topological indices and the presence/absence of Majorana fermions depend on the symmetry of the superconducting gap function and the topology of Fermi surfaces. 
Therefore, a thorough investigation of the nature of the superconductivity and the pairing mechanism is highly desirable.

In early studies, the ferromagnetic quantum critical fluctuation has been indicated~\cite{Ran2019,Sundar2019UTe2,Tokunaga2019} and considered to be a glue of spin-triplet Cooper pairs. On the other hand, the neutron scattering experiments detected antiferromagnetic fluctuation~\cite{duan202incommensurate,Knafo2021,duan2021resonance,raymond2021feedback} with the wave vector around ${\bm q} \simeq (0,\pi,0)$ and called for a search for spin-triplet superconductivity arising from antiferromagnetic fluctuation. To study the relationship between magnetic fluctuation and superconductivity, information on the electron band structure is indispensable~\cite{Kreisel2022}.
An ARPES experiment~\cite{Miao2020ARPES} observed quasi-two-dimensional Fermi surfaces consistent with the LDA+$U$~\cite{Ishizuka2019} and LDA+DMFT calculations~\cite{Xu2019DMFT,Miao2020ARPES} for a large Coulomb interaction. 
Indication of a three-dimensional Fermi surface was also reported~\cite{Miao2020ARPES}, but it is under debate~\cite{fujimori2021}. 
Recent development in crystal growth~\cite{Sakai2022crystal} enabled quantum oscillation measurements~\cite{Aoki2022dHvA,eaton2023quasi2d,broyles2023revealing,aoki2023haasvan}, which precisely detected quasi-two-dimensional Fermi surfaces. The presence or absence of a three-dimensional Fermi surface is still controversial since it is detected only in high fields~\cite{broyles2023revealing}.

Typical uranium-based heavy-fermion superconductors, such as UCoGe and URhGe, exhibit changes in the nature of superconductivity when subjected to pressure~\cite{aokiFMSCreview}.
More interestingly, application of pressure to UTe$_2$ revealed the presence of multiple superconducting phases~\cite{Braithwaite2019,Aoki2020,Lin2020,RanPRB2020,Thomas2020,Knebel2020,aoki2021fieldinduced,Rosa2022}.
This experiment suggests that pressure changes the 
electronic state and superconductivity. 
Therefore, theoretical studies revealing intimate relations between electronic states, magnetic fluctuation, and superconductivity have been awaited.

Inspired by the experimental studies on magnetic fluctuation and band structure in UTe$_2$, we theoretically investigate Fermi surfaces, magnetic fluctuation, and superconductivity in UTe$_2$. Although several microscopic models were analyzed for UTe$_2$~\cite{Ishizuka2020,Shishidou2021,Hazra2023,Kreisel2022,chen2021multiorbital,choi2023correlated,yu2023theory}, a model reproducing antiferromagnetic fluctuation, Fermi surfaces, and spin-triplet superconductivity was not reported and is highly awaited. In this Letter, we propose a mixed-dimensional periodic Anderson model (PAM) consistent with the band structure calculations~\cite{Ishizuka2019,Xu2019DMFT,Miao2020ARPES,Shishidou2021} and reveal antiferromagnetic fluctuation and spin-triplet superconductivity coherently.

\textit{Model and method.} --- 
We construct a mixed-dimensional PAM. With a 
tight-binding model for $f$, $d$, and $p$ electrons as a non-interacting part $H_0$, 
the model is given by $H=H_0+H_{\rm I}$, where 
\begin{eqnarray}
\hspace{-4mm}
&H_0&=H_{f}+H_{d}+H_{p}+H_{fd}+H_{fp}+H_{dp},
\label{eq:PAM}\\
\hspace{-4mm}
    &H_{m}&=\sum_{\bm{k}\sigma}\left(\varepsilon_{m}(\bm{k})-\mu \right)
    a_{m\bm{k}\sigma}^{\dagger}a_{m\bm{k}\sigma}\quad(m=f,d,p),\\
\hspace{-4mm}
    &H_{ml}&=\sum_{\bm{k}\sigma}V_{ml}
    a_{m\bm{k}\sigma}^{\dagger}a_{l\bm{k}\sigma} + {\rm h.c.}
    \quad(ml=fd,fp,dp),
\end{eqnarray}
and $H_{\rm I} = U\sum_{j}n_{f j\uparrow}n_{f j\downarrow}$ is the on-site Coulomb interaction of $f$ electrons. 
The mixed-dimensional property of the model is represented in the kinetic energies, 
\begin{eqnarray}
\varepsilon_{f}(\bm{k})
&=&-2t_{fx}\cos k_x -2t_{fy}\cos k_y\nonumber \\
&&+ 2t_{fz}(\cos k_z +1) +\Delta_f, \\
\varepsilon_{d}(\bm{k})
&=&-2t_{dx}\cos k_x -2t_{dy}\cos k_y +\Delta_d, \\
\varepsilon_{p}(\bm{k})
&=&2t_{px}\cos k_x +2t_{py}\cos k_y +\Delta_p.
\end{eqnarray}
Here, we assume quasi-one-dimensional conduction electrons in accordance with the band structure of UTe$_2$~\cite{Ishizuka2019,Miao2020ARPES,Xu2019DMFT}. 
In UTe$_2$, the $d$ and $p$ electrons mainly conduct along the $a$ and $b$-axis, respectively, corresponding to $t_{dx} \gg t_{dy}$ and $t_{px} \ll t_{py}$.  
We also assume three-dimensional $f$ electrons with comparable hopping integrals $t_{fx}$, $t_{fy}$, and $t_{fz}$. 
Thus, conduction electrons and $f$ electrons have nonequivalent dimensionality, and thus, we call the Hamiltonian the $1$+$3$-dimensional PAM. 
The $1$+$2$-dimensional PAM can also be constructed by simply setting $t_{fz}=0$.  The results of the $1$+$2$-dimensional model are shown in Supplemental Materials for comparison~\cite{Supplemental}.

The crystal fields are denoted by $\Delta_{p,d,f}$, and we adopt momentum-independent hybridization terms for simplicity.
Later, we show that the $f$-electron level $\Delta_f$ and the hybridization with $p$ electrons $V_{fp}$ are control parameters of the model. 
We set $V_{fp}=0.05$ unless mentioned otherwise.
In the following discussions, we fix the other parameters
$(t_{fx},t_{fy},t_{fz},t_{dx},t_{dy},\Delta_d,t_{px},t_{py},\Delta_p, V_{fd},V_{dp},\mu) =
(0.08, 0.035, 0.1, 0.5, 0, 0, 0, 1, 0, 0.05, 0.05, -0.1)$ and the temperature $T=0.01$.

The non-interacting Green function is defined as 
$
    \hat{G}(k)=(i\omega_n \hat{I}-\hat{H}_0)^{-1},
$
where $k=(\bm{k},i\omega_n)$, $\omega_n=(2n+1)\pi T$ are fermionic Matsubata frequencies, and $\hat{I}$ is the identity matrix.
The spin and charge susceptibilities are evaluated by the random phase approximation (RPA) as
\begin{eqnarray}
\begin{split}
    &\chi_s(q)=\frac{\chi^0(q)}{1-U\chi^0(q)}, 
    \quad
    \chi_c(q)=\frac{\chi^0(q)}{1+U\chi^0(q)},
\end{split}
\end{eqnarray}
with the bare susceptibility
\begin{eqnarray}
    \chi^0(q)=-\frac{T}{N}\sum_{k} G_f(k+q)G_f(k).
\end{eqnarray}
Here, $G_f(k)$ is the $f$-electron's Green function, $q=(\bm{q},i\Omega_m)$, and  $\Omega_m=2m\pi T$ are bosonic Matsubata frequencies.
We investigate superconductivity in this model by solving the \'{E}liashberg equation~\cite{yanase2003review} with the effective interactions given by the RPA,
\begin{eqnarray}
\hspace{-4mm}
    &V^s(k-k')&=U+\frac{3}{2}U^2 \chi_s(k-k')-\frac{1}{2}U^2 \chi_c(k-k'),\\
\hspace{-4mm}
    &V^t(k-k')&=-\frac{1}{2}U^2 \chi_s(k-k')-\frac{1}{2}U^2 \chi_c(k-k'),
\end{eqnarray}
where the subscript $s$ and $t$ represent the spin-singlet and spin-triplet Cooper pairing channel, respectively.
The instability of superconductivity is examined by the linearized \'{E}liashberg equation: 
\begin{eqnarray}
    \lambda\Delta(k)=-\frac{T}{N}\sum_{k'} V(k-k') |G_f(k')|^2 \Delta(k').
\end{eqnarray}
The superconducting transition temperature is determined by the condition, $\lambda = 1$.
In the following, we discuss superconducting states by calculating the eigenvalue $\lambda$.
The larger $\lambda$ indicates the higher transition temperature.
The numerical calculations are carried out for the $N=64\times64\times64$ lattice and the Matsubara frequency cutoff $N_f=1024$.

\begin{figure}[htbp]
  \centering  \includegraphics[width=0.48\textwidth]{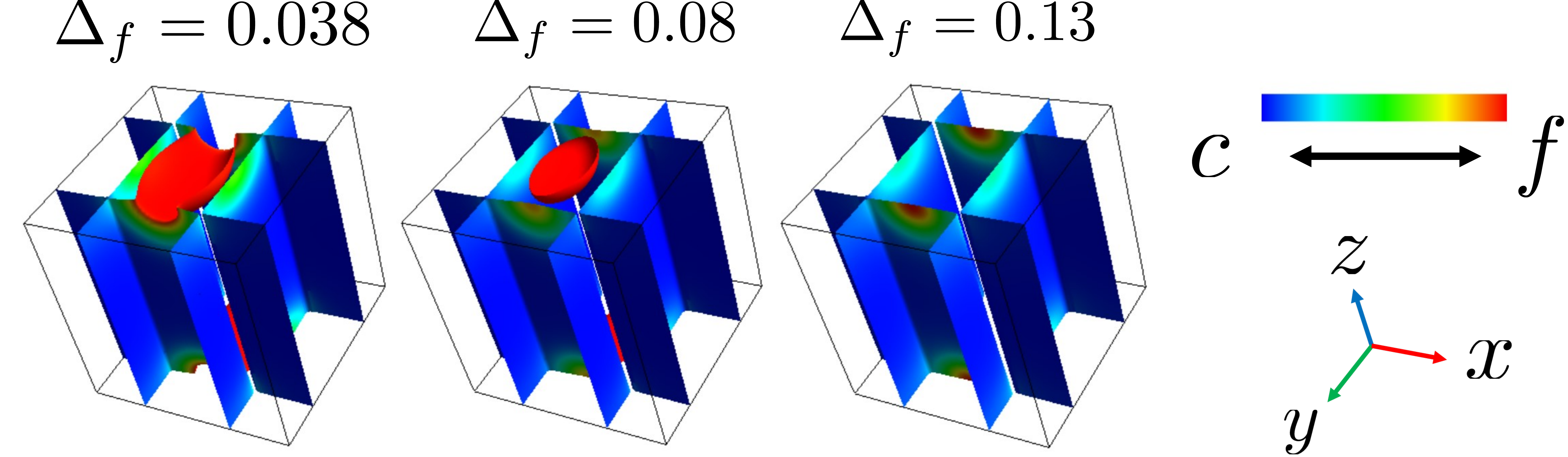}
  \caption{Fermi surfaces of the mixed-dimensional PAM in Eq.~\eqref{eq:PAM}.
  We set $V_{fp}=0.05$ and vary $\Delta_{f}$ from 0.038 to 0.13.
  The weight of $f$ electrons is illustrated in red, while the conduction electrons in blue.
  }
\label{fig:Deltaf_Fermi}
\end{figure}

\textit{Fermi surface.} --- 
First, we show Fermi surfaces of the model.
Later we see that magnetic fluctuation drastically changes depending on the shape of Fermi surfaces, which is mainly determined by $\Delta_f$ and $V_{fp}$.
As shown in Fig.~\ref{fig:Deltaf_Fermi}, 
when we decrease the $f$-electron level $\Delta_f$, 
the Lifshitz transitions successively occur.
When we set $\Delta_f=0.13$, 
two-dimensional rectangular-shaped Fermi surfaces appear as a consequence of the hybridization of one-dimensional $p$ and $d$ electron bands. A considerable weight of $f$ electrons also exists on the two-dimensional Fermi surfaces owing to the $c$-$f$ hybridization.
A small three-dimensional Fermi surface is present when we decrease the $f$-electron level as $\Delta_f=0.08$.
By further decreasing $\Delta_f$, the three-dimensional Fermi surface is expanded and connected with the two-dimensional Fermi surfaces (see Fig.~\ref{fig:Deltaf_Fermi} for $\Delta_f=0.038$).
A similar change in Fermi surfaces was reported in the band structure calculations for UTe$_2$. 
The LDA+$U$ and LDA+DMFT calculations for a large Coulomb interaction concluded two-dimensional Fermi surfaces similar to our model for $\Delta_f=0.13$~\cite{Miao2020ARPES,Ishizuka2019,Xu2019DMFT}, while the LDA+$U$ calculation for an intermediate Coulomb interaction obtained mixed-dimensional Fermi surfaces as reproduced in our model for $\Delta_f=0.038$~\cite{Ishizuka2019}. 
Thus, the parameter $\Delta_f$ in our model may correspond to the interaction parameter of UTe$_2$.

\begin{figure}[htbp]
    \centering
\includegraphics[width=0.4\textwidth]{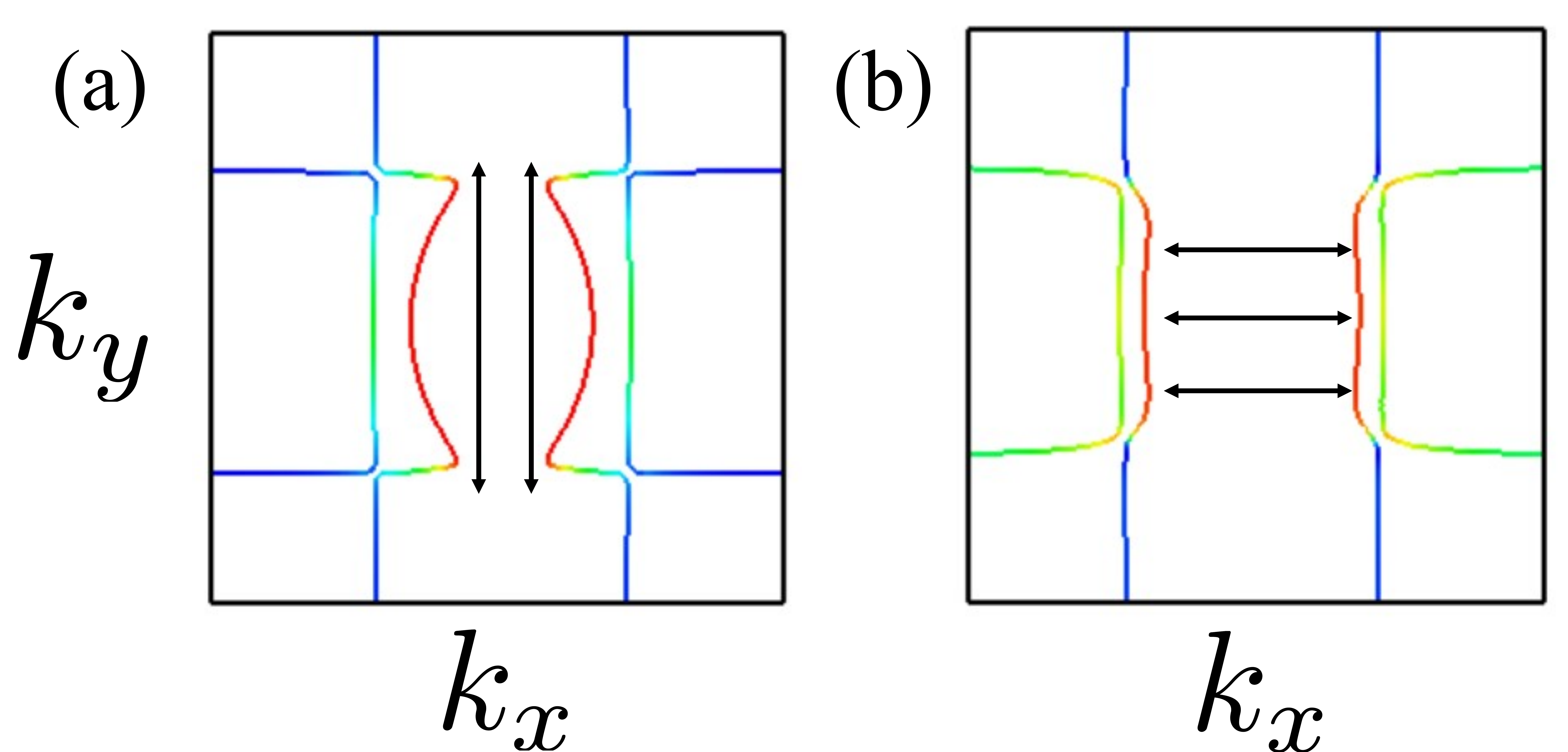}
    \caption{
    Fermi surfaces on the $k_z=\pi$ plane for $\Delta_f=0.05$. 
    (a) $V_{fp}=0.05$ and (b) $V_{fp}=0.26$.
    The weight of $f$ electrons is shown by color as in Fig.~\ref{fig:Deltaf_Fermi}, and nesting vectors are illustrated by arrows.
    }
    \label{fig:nesting}
\end{figure}

In heavy-fermion systems, quantum phase transitions are often tunable by the applied pressure, which has been interpreted as an increase in hybridization. 
The following results in our model are also sensitive to the hybridization parameter $V_{fp}$, which affects the shape of the three-dimensional Fermi surface rather than its size. 
We show the Fermi surfaces for various $V_{fp}$ in Fig.~\ref{fig:nesting}. Interestingly, the nesting property changes with $V_{fp}$, as illustrated in the figure.

\begin{figure}[htbp]
    \centering
\includegraphics[width=0.48\textwidth]{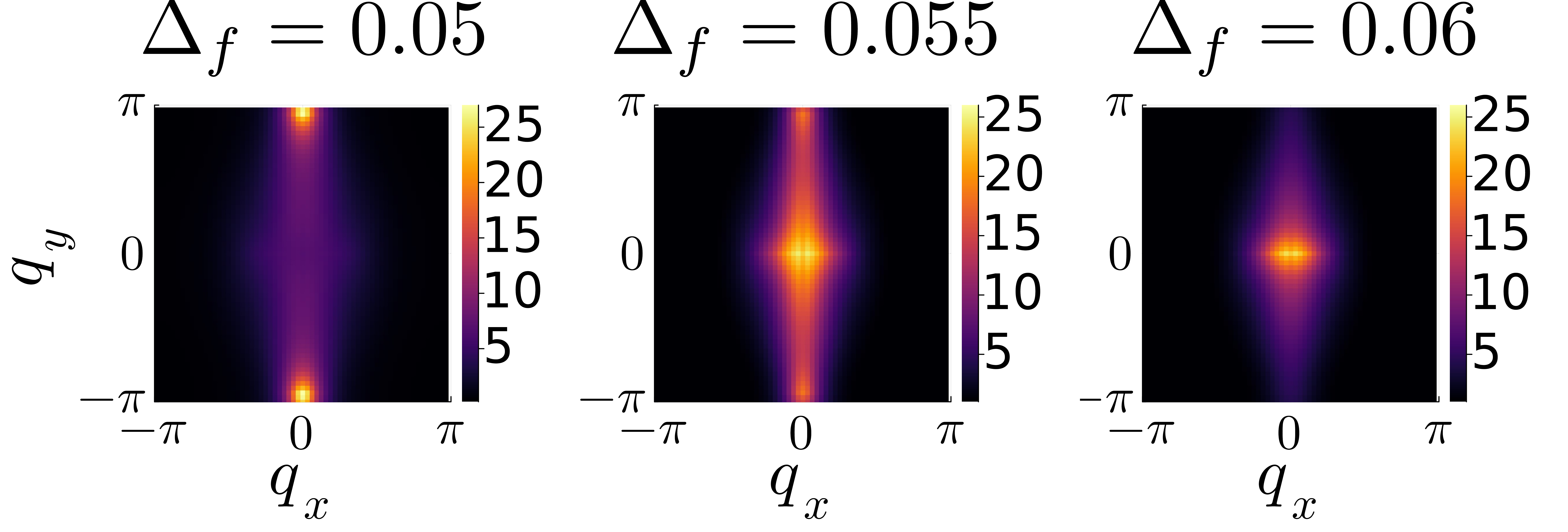}
    \caption{
    Static spin susceptibility $\chi_s(\bm q) \equiv \chi_s(\bm q,i\Omega_m=0)$ on the $q_z=0$ plane for various $f$-electron levels, $\Delta_f$.
    }
\label{fig:Deltaf_susc}
\end{figure}

\textit{Spin susceptibility.} --- 
Second, we discuss magnetic fluctuation. Our mixed-dimensional PAM shows ferromagnetic fluctuation, antiferromagnetic fluctuation, and their coexistence depending on the parameters discussed above. 
Antiferromagnetic fluctuation is dominant for the $f$-electron level $\Delta_f$ ranging from 0.038 to 0.054, while ferromagnetic fluctuation is dominant for $\Delta_f$ from 0.055 to 0.13.
To focus on the parameter dependence, we set the Coulomb interaction so that the Stoner factor is ${\rm max} U \chi^0(q)=0.98$.
Magnetic susceptibility on the $q_z=0$ plane is shown in Fig.~\ref{fig:Deltaf_susc}.

In the result for $\Delta_f=0.05$, we see antiferromagnetic fluctuation with the wave vector around $\bm{q}=(0,\pi,0)$, consistent with the neutron scattering experiments~\cite{duan202incommensurate,Knafo2021}.
Thus, antiferromagnetic fluctuation in UTe${}_2$ can be explained by the nesting vector of Fermi surfaces with substantial $f$-electrons, as shown in Fig.~\ref{fig:nesting}(a).
On the other hand, when we slightly increase the $f$-electron level as $\Delta_f=0.055$, ferromagnetic fluctuation with the wave vector $\bm{q}=0$ becomes dominant and coexists with antiferromagnetic fluctuation.
Ferromagnetic fluctuation naturally arises from the three-dimensional property of $f$ electrons.
Thus, (anti)ferromagnetic fluctuations are sensitive to the parameters, and they can coexist. 
This finding may be consistent with the fact that UTe${}_2$ has been considered to be a nearly ferromagnetic system~\cite{Ran2019,Sundar2019UTe2,Tokunaga2019,fujibayashi2023,Aoki2022review} and coexisting ferro- and antiferromagnetic fluctuations were reported~\cite{ambika2022,Xu2019DMFT}.

\begin{figure}[htbp]
    \centering
    \includegraphics[width=0.48\textwidth]{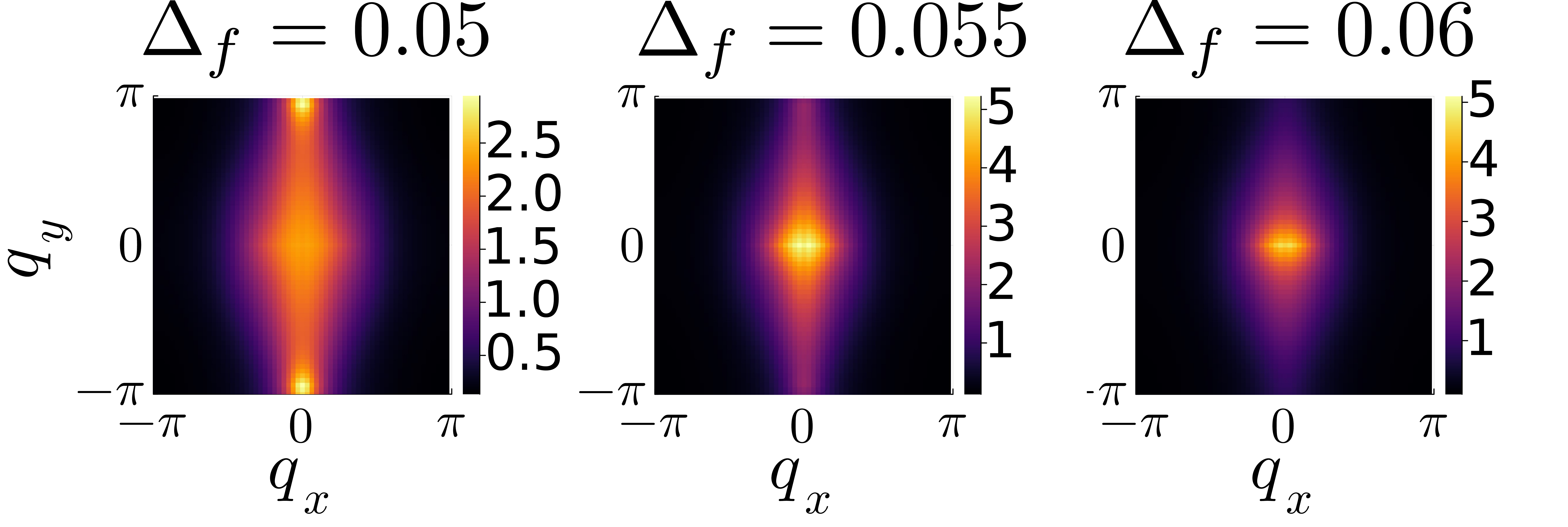}
    \caption{
    Spin susceptibility integrated along the $q_z$ direction, defined by $\chi_z^{\rm 2d}(q_x,q_y) \equiv \int \chi_s(\bm q) {\rm d}q_z$.
    }
    \label{fig:Deltaf_int}
\end{figure}

In our model, magnetic susceptibility always shows maximum on the $q_z=0$ plane. To get information on magnetic fluctuation away from $q_z=0$, Fig.~\ref{fig:Deltaf_int} shows the integration of spin susceptibility with respect to $q_z$. 
Even for $\Delta_f = 0.05$ resulting in dominant antiferromagnetic fluctuation, magnetic fluctuation has substantial weight around $(q_x,q_y)=(0,0)$, indicating two-dimensional ferromagnetic fluctuation.
Because two-dimensional fluctuation generally favors superconductivity~\cite{yanase2003review}, ferromagnetic correlation is expected to play an important role in superconductivity.

\begin{figure}[htbp]
    \centering
\includegraphics[width=0.48\textwidth]{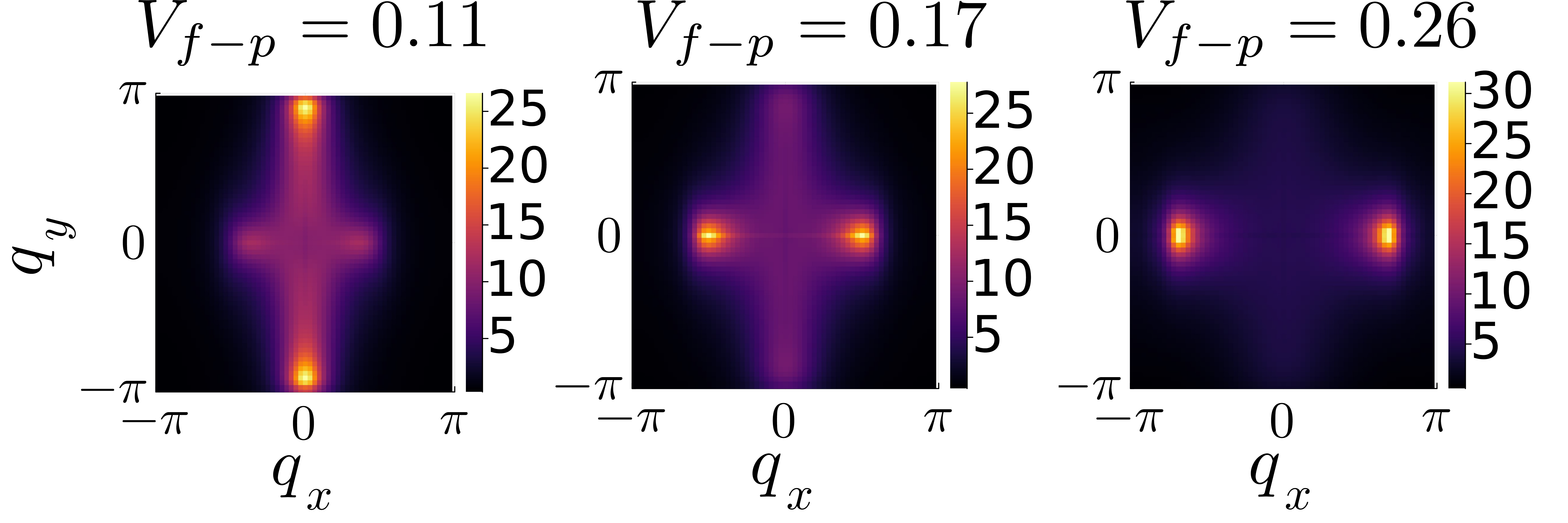}
    \caption{
    Spin susceptibility on the $q_z=0$ plane for various hybridization parameters, $V_{fp}$.
    We fix the $f$-electron level $\Delta_f = 0.05$.
    }
    \label{fig:Vfp_susc}
\end{figure}

As shown above, the mixed-dimensional PAM reproduces both ferromagnetic and antiferromagnetic fluctuations that are considered to exist in UTe${}_2$. 
The three-dimensional $f$ electrons are crucially important for magnetic fluctuations~\cite{Supplemental}.
Furthermore, the model can reproduce a theoretical study for UTe$_2$ under pressures~\cite{Ishizuka2020}, in which antiferromagnetic fluctuation with the wave vector ${\bm q} \simeq (\pi,0,0)$ develops by increasing $c$-$f$ hybridization. 
Figure~\ref{fig:Vfp_susc} shows the spin susceptibility in the mixed-dimensional PAM for strong hybridization. 
We see that the wave vector of antiferromagnetic fluctuation changes from ${\bm q} \simeq (0,\pi,0)$ to ${\bm q} \simeq (\pi,0,0)$.
This is because the strong hybridization changes the Fermi surfaces and results in better nesting along the $x$-direction as shown in Fig.~\ref{fig:nesting}(b).

\textit{Superconductivity.} --- 
Next, we examine superconductivity mediated by magnetic fluctuations. In particular, we focus on the relation between the wave vector of magnetic fluctuation and the symmetry of superconductivity. 

\begin{table}[hbtp]
  \caption{Irreducible representations and basis functions of the $D_{2h}$ point group.}
  \label{table:representation}
  \centering
  \begin{tabular}{|c|c|c|c|c|c|c|c|c|}
    \hline
    IR($D_{2h}$) & $A_g$ & $B_{1g}$ & $B_{2g}$ & $B_{3g}$ & $A_u$ & $B_{1u}$ & $B_{2u}$ & $B_{3u}$ \\
    \hline
    Basis function & $1$ & $k_x k_y$ & $k_x k_z$ & $k_y k_z$  & $k_x k_y k_z$ & $k_z$ & $k_y$ & $k_x$\\
    \hline
  \end{tabular}
\end{table}

For this purpose, we solve the linearized \'{E}liashberg equation for all the irreducible representations of the $D_{2h}$ point group. 
Note that the model in this study does not include a spin-orbit coupling. Therefore, we classify the gap function $\Delta(k)$ by the  representation without a spin degree of freedom.
Table~\ref{table:representation} shows the list of irreducible representations and basis functions. In this classification, the $B_{1u}$, $B_{2u}$, and $B_{3u}$ representations correspond to the $p$-wave superconductivity while the $A_{u}$ representation does the $f$-wave superconductivity. The other representations indicate spin-singlet superconductivity with either $s$-wave or $d$-wave symmetry.

\begin{figure}[htbp]
    \centering
\includegraphics[width=0.45\textwidth]{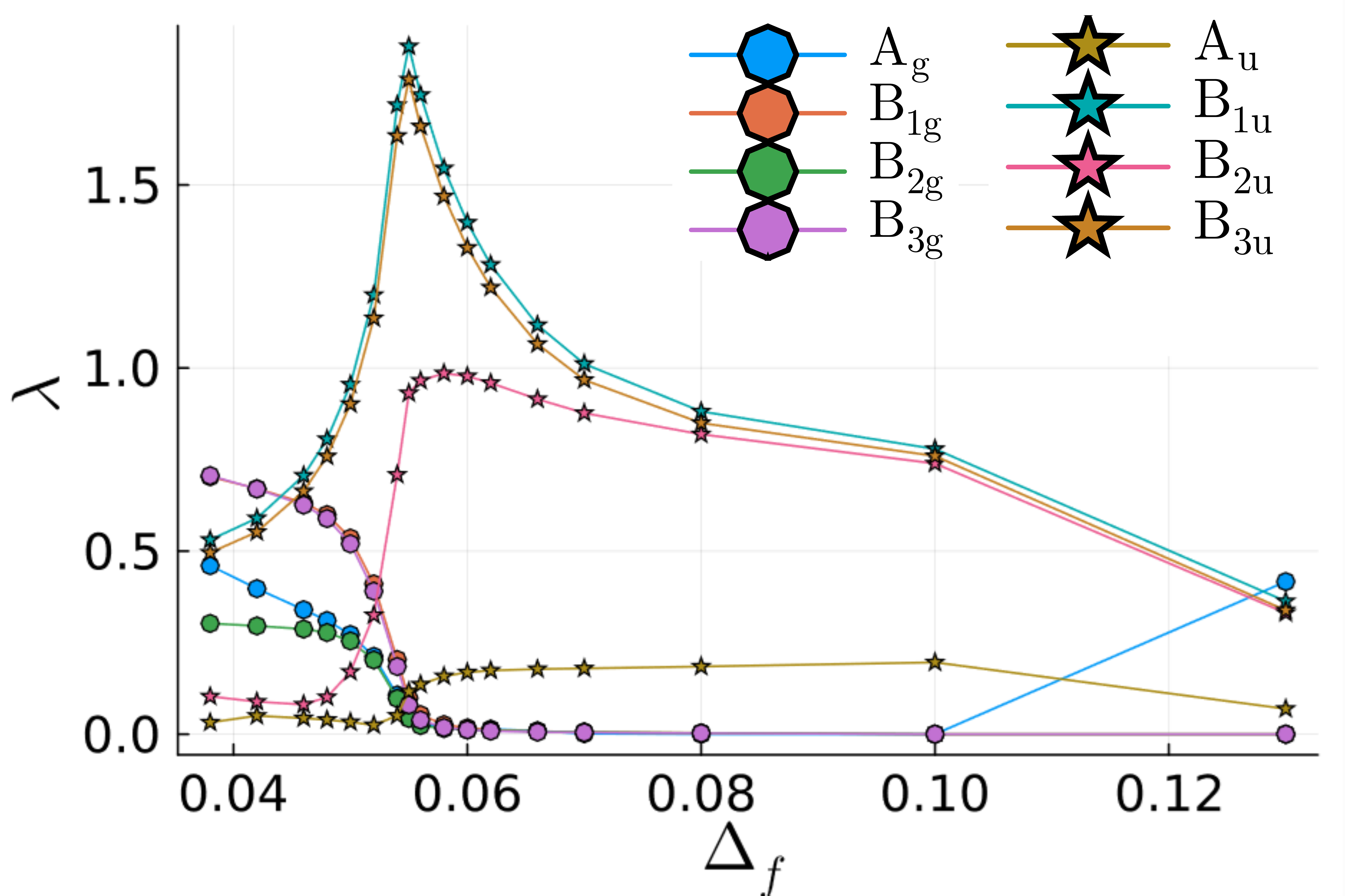}
    \caption{
    Eigenvalues of \'{E}liashberg equation as a function of the $f$-electron level $\Delta_f$. The maximum values for each irreducible representation are shown. Spin-triplet superconductivity and spin-singlet superconductivity are plotted by stars and circles, respectively.
    }
    \label{fig:Deltaf_dep}
\end{figure}

First, we show Fig.~\ref{fig:Deltaf_dep} for the 
$\Delta_f$ dependence of eigenvalues of \'{E}liashberg equation. 
We see that spin-triplet $p$-wave superconductivity of $B_{1u}$ and $B_{3u}$ representations is stable over a wide range of parameters.
In particular, the $p$-wave superconductivity is stable even around $\Delta_f=0.05$ where 
antiferromagnetic fluctuation is dominant. This is because ferromagnetic fluctuation and antiferromagnetic fluctuation cooperatively stabilize the $p$-wave superconductivity. Indeed, the $p$-wave superconductivity is most stable when the two fluctuations coexist, as the eigenvalue $\lambda$ shows the maximum around $\Delta_f=0.055$. 

\begin{figure}[htbp]
    \centering
\includegraphics[width=0.4\textwidth]{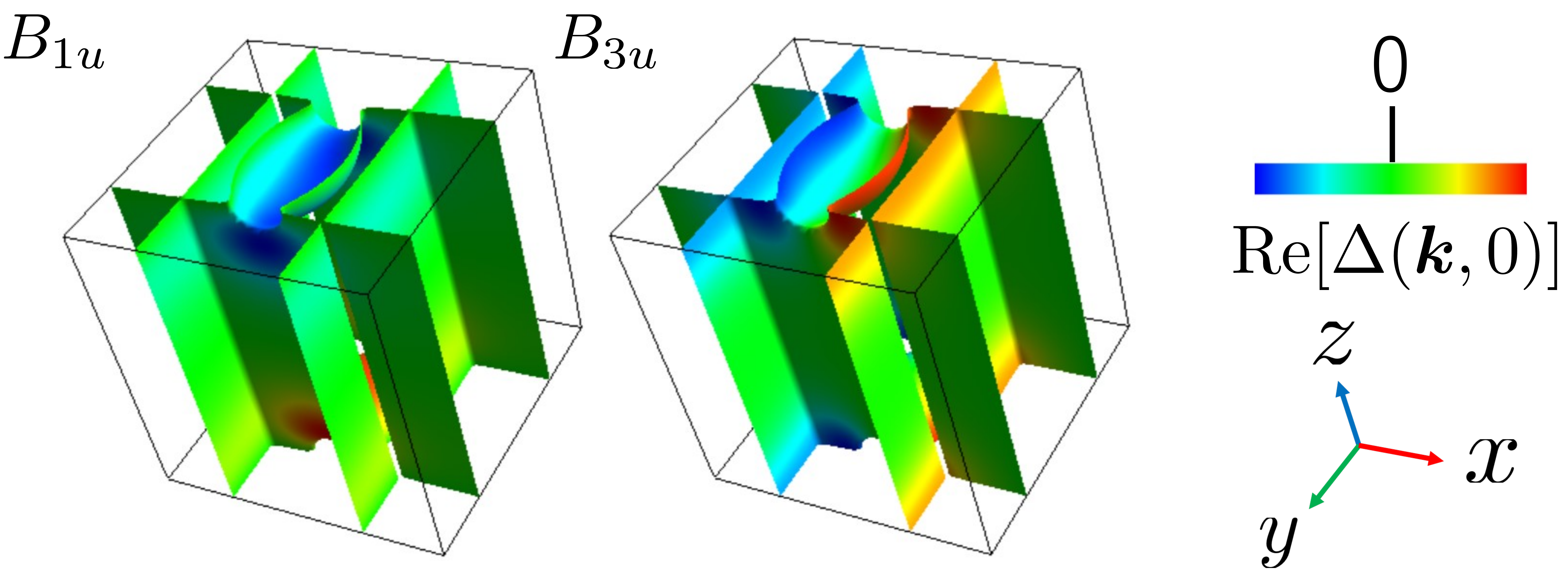}
    \caption{
    The gap functions of $B_{1u}$ and $B_{3u}$ representations on the Fermi surfaces. The color represents the real part of the gap function.
    }
    \label{fig:Vfp}
\end{figure}

The antiferromagnetic fluctuation can favor spin-triplet superconductivity when the gap function has the same sign between the Fermi momentum connected by the nesting vector~\cite{yanase2003review}. This condition is indeed satisfied for the $B_{1u}$ and $B_{3u}$ representations. 
Figure~\ref{fig:Vfp} shows the gap functions on the Fermi surfaces for $\Delta_f=0.05$, where the static gap function is defined as 
\begin{eqnarray}
    \Delta(\bm{k},0) \equiv \frac{\Delta(\bm{k},i\pi T)+\Delta(\bm{k},-i\pi T)}{2}.
\end{eqnarray}
Antiferromagnetic fluctuation with $\bm{q}=(0,\pi,0)$ acts as an attractive force for the $B_{1u}$ and $B_{3u}$ representations because the gap functions on the Fermi surface do not change the sign for the magnetic scattering. However, it gives a repulsive interaction for the $B_{2u}$
representation, and thus the $B_{2u}$ state is suppressed. 
For $\Delta_f >0.08$, all the $p$-wave superconducting states are nearly degenerate, because the antiferromagnetic fluctuation is negligible.

\begin{figure}[htbp]
    \centering
\includegraphics[width=0.4\textwidth]{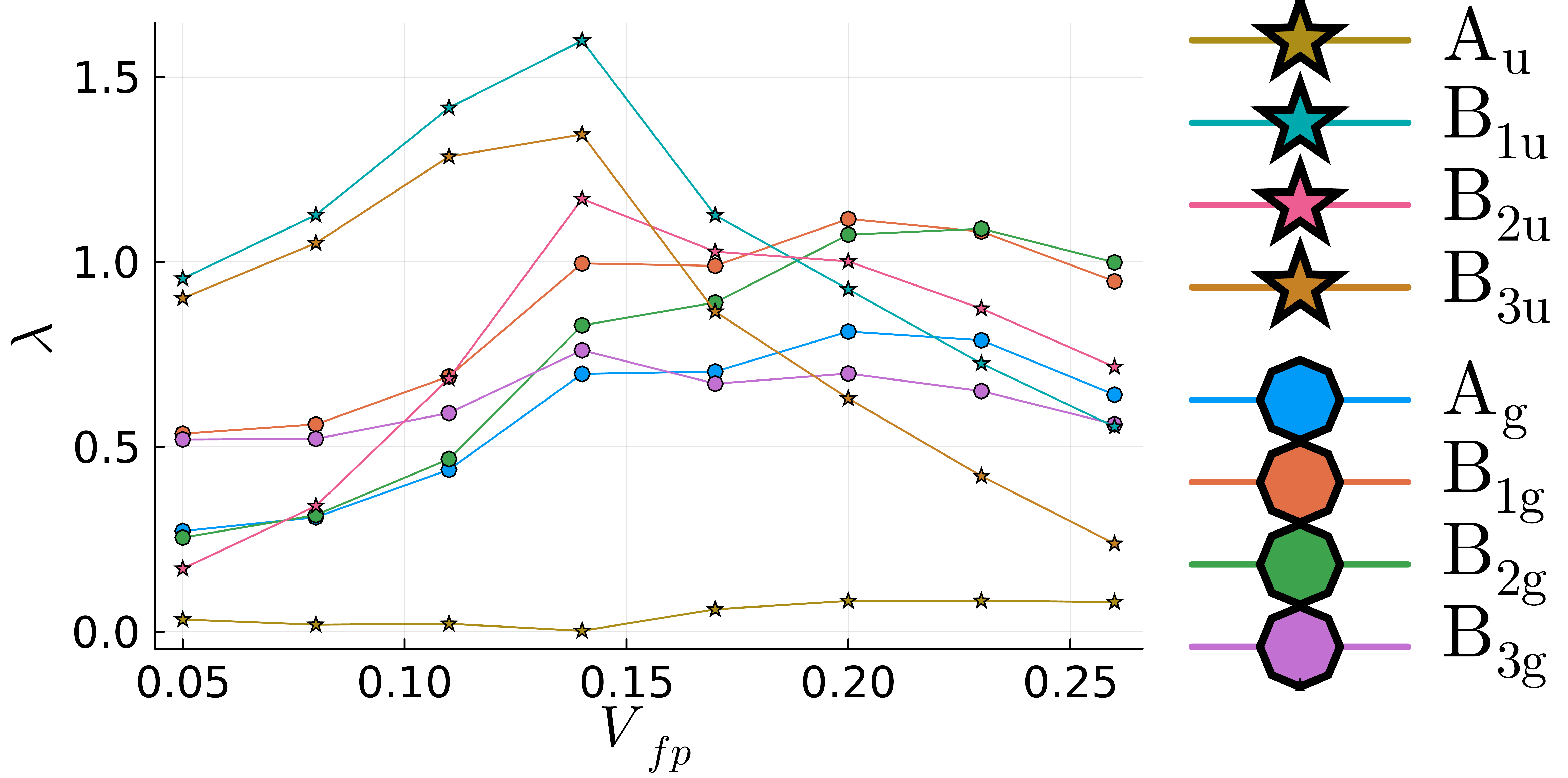}
    \caption{
    The hybridization dependence of superconductivity, indicated by the eigenvalues $\lambda$ of \'{E}liashberg equation. We set $\Delta_{f}=0.05$.
    }
    \label{fig:Vfp_dep}
\end{figure}

Next, we show the hybridization $V_{fp}$ dependence of superconductivity. 
In Fig.~\ref{fig:Vfp_dep}, 
we see that stable superconducting states are the spin-triplet $B_{1u}$ and $B_{3u}$ states for small $V_{fp}$, while the spin-singlet $B_{1g}$ and $B_{2g}$ states are stable for large $V_{fp}$. 
This can be regarded as a parity transition of superconductivity, where the symmetry of superconductivity changes from odd-parity to even-parity by increasing the magnitude of hybridization. Accompanied by the development of antiferromagnetic fluctuation with ${\bm q} \parallel \hat{x}$, the spin-singlet superconductivity of the $B_{1g}$ and $B_{2g}$ representations is favored. 
The hybridization dependence obtained above is qualitatively consistent with the previous study~\cite{Ishizuka2020}, where the extended $s$-wave state is stable instead of the $d$-wave one.
Since the hybridization is expected to be enhanced by pressure in heavy-fermion systems, the parity transition may correspond to the multiple superconducting phases of UTe$_2$ under pressures~\cite{Braithwaite2019,Aoki2022review}. However, the pressure changes various parameters including the $f$-electron level and Coulomb interaction, and therefore, further verification based on first-principles calculations combined with experiments is desirable.

\textit{Summary and discussion.} ---
In this Letter, we have proposed the $1$+$3$-dimensional PAM as a minimal model for Fermi surfaces, magnetic fluctuation, and superconductivity in UTe$_2$. The model reproduces the antiferromagnetic and ferromagnetic fluctuations which have been indicated by experiments. The spin-triplet superconductivity is cooperatively stabilized by these magnetic fluctuations. We emphasize that the spin-triplet $p$-wave superconductivity can coexist with antiferromagnetic fluctuation in contrast to the common belief. This is a characteristic property of the $1$+$3$-dimensional model, as we find that spin-triplet superconductivity and antiferromagnetic fluctuation are exclusive in the $1$+$2$-dimensional PAM~\cite{Supplemental}. 

We can extend the mixed-dimensional PAM to include spin-orbit coupling, magnetic field, and multiple orbitals and sublattices. Thus, we expect that the model will be the basis of further theoretical studies for UTe$_2$. Our calculation indicates the $k_x$ and $k_z$ orbital components of order parameters. However, the spin-triplet superconducting states are degenerate because of the spin component in our model. Spin-orbit coupling and magnetic fields break the spin SU(2) symmetry and play essential roles in determining the spin component.  
Thus, calculations containing these effects are desirable in the future to identify the multiple superconducting phases under pressures~\cite{Braithwaite2019,Aoki2020,Lin2020,RanPRB2020,Thomas2020,Knebel2020,aoki2021fieldinduced,Rosa2022,Aoki2022review} and/or magnetic fields~\cite{ran2019extreme,Rosuel2023,Kinjo2023,tokiwa2022stabilization,sakai2022field,Aoki2022review}. 
Recently, development in the high-purity crystal growth~\cite{Sakai2022crystal} makes a second era in experiments of UTe$_2$. 
For instance, the quantum oscillation was detected~\cite{Aoki2022dHvA,eaton2023quasi2d,broyles2023revealing,aoki2023haasvan}, and the thermal conductivity~\cite{Suetsugu2023} and NMR Knight shift~\cite{matsumura2023large} show behaviors consistent with the order parameter ${\bm d}=k_x \hat{x} +k_y \hat{y} + k_z \hat{z}$, similar to the B-phase of $^3$He~\cite{Leggett1975}. The order parameter contains the $k_x$ and $k_z$ orbital components consistent with our results and realizes strong (weak) topological superconductivity in the presence (absence) of a three-dimensional Fermi surface~\cite{Ishizuka2019,Tei2023}.

\begin{acknowledgments}
We appreciate helpful discussions with J. Ishizuka, D. Aoki, J.-P. Brison, K. Ishida, G. Knebel, Y. Matsuda, S. Suetsugu, Y. Tokiwa, and Y. Tokunaga.
We use FermiSurfer to describe the Fermi surfaces in this paper~\cite{kawamura2019fermisurfer}.
This work was supported by
JSPS KAKENHI (Grant Nos. JP21K18145, JP22H01181, JP22H04933). 
\end{acknowledgments}

\bibliography{ref}

\providecommand{\noopsort}[1]{}\providecommand{\singleletter}[1]{#1}%
\begin{thebibliography}{52}%
\makeatletter
\providecommand \@ifxundefined [1]{%
 \@ifx{#1\undefined}
}%
\providecommand \@ifnum [1]{%
 \ifnum #1\expandafter \@firstoftwo
 \else \expandafter \@secondoftwo
 \fi
}%
\providecommand \@ifx [1]{%
 \ifx #1\expandafter \@firstoftwo
 \else \expandafter \@secondoftwo
 \fi
}%
\providecommand \natexlab [1]{#1}%
\providecommand \enquote  [1]{``#1''}%
\providecommand \bibnamefont  [1]{#1}%
\providecommand \bibfnamefont [1]{#1}%
\providecommand \citenamefont [1]{#1}%
\providecommand \href@noop [0]{\@secondoftwo}%
\providecommand \href [0]{\begingroup \@sanitize@url \@href}%
\providecommand \@href[1]{\@@startlink{#1}\@@href}%
\providecommand \@@href[1]{\endgroup#1\@@endlink}%
\providecommand \@sanitize@url [0]{\catcode `\\12\catcode `\$12\catcode
  `\&12\catcode `\#12\catcode `\^12\catcode `\_12\catcode `\%12\relax}%
\providecommand \@@startlink[1]{}%
\providecommand \@@endlink[0]{}%
\providecommand \url  [0]{\begingroup\@sanitize@url \@url }%
\providecommand \@url [1]{\endgroup\@href {#1}{\urlprefix }}%
\providecommand \urlprefix  [0]{URL }%
\providecommand \Eprint [0]{\href }%
\providecommand \doibase [0]{http://dx.doi.org/}%
\providecommand \selectlanguage [0]{\@gobble}%
\providecommand \bibinfo  [0]{\@secondoftwo}%
\providecommand \bibfield  [0]{\@secondoftwo}%
\providecommand \translation [1]{[#1]}%
\providecommand \BibitemOpen [0]{}%
\providecommand \bibitemStop [0]{}%
\providecommand \bibitemNoStop [0]{.\EOS\space}%
\providecommand \EOS [0]{\spacefactor3000\relax}%
\providecommand \BibitemShut  [1]{\csname bibitem#1\endcsname}%
\let\auto@bib@innerbib\@empty
\bibitem [{\citenamefont {Schnyder}\ \emph {et~al.}(2008)\citenamefont
  {Schnyder}, \citenamefont {Ryu}, \citenamefont {Furusaki},\ and\
  \citenamefont {Ludwig}}]{Schnyder2008}%
  \BibitemOpen
  \bibfield  {author} {\bibinfo {author} {\bibfnamefont {A.~P.}\ \bibnamefont
  {Schnyder}}, \bibinfo {author} {\bibfnamefont {S.}~\bibnamefont {Ryu}},
  \bibinfo {author} {\bibfnamefont {A.}~\bibnamefont {Furusaki}}, \ and\
  \bibinfo {author} {\bibfnamefont {A.~W.~W.}\ \bibnamefont {Ludwig}},\ }\href
  {\doibase 10.1103/PhysRevB.78.195125} {\bibfield  {journal} {\bibinfo
  {journal} {Phys. Rev. B}\ }\textbf {\bibinfo {volume} {78}},\ \bibinfo
  {pages} {195125} (\bibinfo {year} {2008})}\BibitemShut {NoStop}%
\bibitem [{\citenamefont {Sato}(2010)}]{Sato2010}%
  \BibitemOpen
  \bibfield  {author} {\bibinfo {author} {\bibfnamefont {M.}~\bibnamefont
  {Sato}},\ }\href {\doibase 10.1103/PhysRevB.81.220504} {\bibfield  {journal}
  {\bibinfo  {journal} {Phys. Rev. B}\ }\textbf {\bibinfo {volume} {81}},\
  \bibinfo {pages} {220504} (\bibinfo {year} {2010})}\BibitemShut {NoStop}%
\bibitem [{\citenamefont {Fu}\ and\ \citenamefont {Berg}(2010)}]{Fu-Berg2010}%
  \BibitemOpen
  \bibfield  {author} {\bibinfo {author} {\bibfnamefont {L.}~\bibnamefont
  {Fu}}\ and\ \bibinfo {author} {\bibfnamefont {E.}~\bibnamefont {Berg}},\
  }\href {\doibase 10.1103/PhysRevLett.105.097001} {\bibfield  {journal}
  {\bibinfo  {journal} {Phys. Rev. Lett.}\ }\textbf {\bibinfo {volume} {105}},\
  \bibinfo {pages} {097001} (\bibinfo {year} {2010})}\BibitemShut {NoStop}%
\bibitem [{\citenamefont {Sato}\ and\ \citenamefont
  {Fujimoto}(2016)}]{sato-fujimotoreview}%
  \BibitemOpen
  \bibfield  {author} {\bibinfo {author} {\bibfnamefont {M.}~\bibnamefont
  {Sato}}\ and\ \bibinfo {author} {\bibfnamefont {S.}~\bibnamefont
  {Fujimoto}},\ }\href {\doibase 10.7566/JPSJ.85.072001} {\bibfield  {journal}
  {\bibinfo  {journal} {Journal of the Physical Society of Japan}\ }\textbf
  {\bibinfo {volume} {85}},\ \bibinfo {pages} {072001} (\bibinfo {year}
  {2016})},\ \Eprint
  {http://arxiv.org/abs/https://doi.org/10.7566/JPSJ.85.072001}
  {https://doi.org/10.7566/JPSJ.85.072001} \BibitemShut {NoStop}%
\bibitem [{\citenamefont {Linder}\ and\ \citenamefont
  {Robinson}(2015)}]{Linder2015}%
  \BibitemOpen
  \bibfield  {author} {\bibinfo {author} {\bibfnamefont {J.}~\bibnamefont
  {Linder}}\ and\ \bibinfo {author} {\bibfnamefont {J.~W.~A.}\ \bibnamefont
  {Robinson}},\ }\href {\doibase 10.1038/nphys3242} {\bibfield  {journal}
  {\bibinfo  {journal} {Nature Physics}\ }\textbf {\bibinfo {volume} {11}},\
  \bibinfo {pages} {307} (\bibinfo {year} {2015})}\BibitemShut {NoStop}%
\bibitem [{\citenamefont {Ran}\ \emph {et~al.}(2019{\natexlab{a}})\citenamefont
  {Ran}, \citenamefont {Eckberg}, \citenamefont {Ding}, \citenamefont
  {Furukawa}, \citenamefont {Metz}, \citenamefont {Saha}, \citenamefont {Liu},
  \citenamefont {Zic}, \citenamefont {Kim}, \citenamefont {Paglione},\ and\
  \citenamefont {Butch}}]{Ran2019}%
  \BibitemOpen
  \bibfield  {author} {\bibinfo {author} {\bibfnamefont {S.}~\bibnamefont
  {Ran}}, \bibinfo {author} {\bibfnamefont {C.}~\bibnamefont {Eckberg}},
  \bibinfo {author} {\bibfnamefont {Q.~P.}\ \bibnamefont {Ding}}, \bibinfo
  {author} {\bibfnamefont {Y.}~\bibnamefont {Furukawa}}, \bibinfo {author}
  {\bibfnamefont {T.}~\bibnamefont {Metz}}, \bibinfo {author} {\bibfnamefont
  {S.~R.}\ \bibnamefont {Saha}}, \bibinfo {author} {\bibfnamefont {I.~L.}\
  \bibnamefont {Liu}}, \bibinfo {author} {\bibfnamefont {M.}~\bibnamefont
  {Zic}}, \bibinfo {author} {\bibfnamefont {H.}~\bibnamefont {Kim}}, \bibinfo
  {author} {\bibfnamefont {J.}~\bibnamefont {Paglione}}, \ and\ \bibinfo
  {author} {\bibfnamefont {N.~P.}\ \bibnamefont {Butch}},\ }\href {\doibase
  10.1126/science.aav8645} {\bibfield  {journal} {\bibinfo  {journal}
  {Science}\ }\textbf {\bibinfo {volume} {365}},\ \bibinfo {pages} {684}
  (\bibinfo {year} {2019}{\natexlab{a}})}\BibitemShut {NoStop}%
\bibitem [{\citenamefont {Aoki}\ \emph
  {et~al.}(2022{\natexlab{a}})\citenamefont {Aoki}, \citenamefont {Brison},
  \citenamefont {Flouquet}, \citenamefont {Ishida}, \citenamefont {Knebel},
  \citenamefont {Tokunaga},\ and\ \citenamefont {Yanase}}]{Aoki2022review}%
  \BibitemOpen
  \bibfield  {author} {\bibinfo {author} {\bibfnamefont {D.}~\bibnamefont
  {Aoki}}, \bibinfo {author} {\bibfnamefont {J.-P.}\ \bibnamefont {Brison}},
  \bibinfo {author} {\bibfnamefont {J.}~\bibnamefont {Flouquet}}, \bibinfo
  {author} {\bibfnamefont {K.}~\bibnamefont {Ishida}}, \bibinfo {author}
  {\bibfnamefont {G.}~\bibnamefont {Knebel}}, \bibinfo {author} {\bibfnamefont
  {Y.}~\bibnamefont {Tokunaga}}, \ and\ \bibinfo {author} {\bibfnamefont
  {Y.}~\bibnamefont {Yanase}},\ }\href {\doibase 10.1088/1361-648X/ac5863}
  {\bibfield  {journal} {\bibinfo  {journal} {Journal of Physics: Condensed
  Matter}\ }\textbf {\bibinfo {volume} {34}},\ \bibinfo {pages} {243002}
  (\bibinfo {year} {2022}{\natexlab{a}})}\BibitemShut {NoStop}%
\bibitem [{\citenamefont {Ishizuka}\ \emph {et~al.}(2019)\citenamefont
  {Ishizuka}, \citenamefont {Sumita}, \citenamefont {Daido},\ and\
  \citenamefont {Yanase}}]{Ishizuka2019}%
  \BibitemOpen
  \bibfield  {author} {\bibinfo {author} {\bibfnamefont {J.}~\bibnamefont
  {Ishizuka}}, \bibinfo {author} {\bibfnamefont {S.}~\bibnamefont {Sumita}},
  \bibinfo {author} {\bibfnamefont {A.}~\bibnamefont {Daido}}, \ and\ \bibinfo
  {author} {\bibfnamefont {Y.}~\bibnamefont {Yanase}},\ }\href {\doibase
  10.1103/PhysRevLett.123.217001} {\bibfield  {journal} {\bibinfo  {journal}
  {Phys. Rev. Lett.}\ }\textbf {\bibinfo {volume} {123}},\ \bibinfo {pages}
  {217001} (\bibinfo {year} {2019})}\BibitemShut {NoStop}%
\bibitem [{\citenamefont {Sundar}\ \emph {et~al.}(2019)\citenamefont {Sundar},
  \citenamefont {Gheidi}, \citenamefont {Akintola}, \citenamefont {C\^ot\'e},
  \citenamefont {Dunsiger}, \citenamefont {Ran}, \citenamefont {Butch},
  \citenamefont {Saha}, \citenamefont {Paglione},\ and\ \citenamefont
  {Sonier}}]{Sundar2019UTe2}%
  \BibitemOpen
  \bibfield  {author} {\bibinfo {author} {\bibfnamefont {S.}~\bibnamefont
  {Sundar}}, \bibinfo {author} {\bibfnamefont {S.}~\bibnamefont {Gheidi}},
  \bibinfo {author} {\bibfnamefont {K.}~\bibnamefont {Akintola}}, \bibinfo
  {author} {\bibfnamefont {A.~M.}\ \bibnamefont {C\^ot\'e}}, \bibinfo {author}
  {\bibfnamefont {S.~R.}\ \bibnamefont {Dunsiger}}, \bibinfo {author}
  {\bibfnamefont {S.}~\bibnamefont {Ran}}, \bibinfo {author} {\bibfnamefont
  {N.~P.}\ \bibnamefont {Butch}}, \bibinfo {author} {\bibfnamefont {S.~R.}\
  \bibnamefont {Saha}}, \bibinfo {author} {\bibfnamefont {J.}~\bibnamefont
  {Paglione}}, \ and\ \bibinfo {author} {\bibfnamefont {J.~E.}\ \bibnamefont
  {Sonier}},\ }\href {\doibase 10.1103/PhysRevB.100.140502} {\bibfield
  {journal} {\bibinfo  {journal} {Phys. Rev. B}\ }\textbf {\bibinfo {volume}
  {100}},\ \bibinfo {pages} {140502(R)} (\bibinfo {year} {2019})}\BibitemShut
  {NoStop}%
\bibitem [{\citenamefont {Tokunaga}\ \emph {et~al.}(2019)\citenamefont
  {Tokunaga}, \citenamefont {Sakai}, \citenamefont {Kambe}, \citenamefont
  {Hattori}, \citenamefont {Higa}, \citenamefont {Nakamine}, \citenamefont
  {Kitagawa}, \citenamefont {Ishida}, \citenamefont {Nakamura}, \citenamefont
  {Shimizu}, \citenamefont {Homma}, \citenamefont {Li}, \citenamefont {Honda},\
  and\ \citenamefont {Aoki}}]{Tokunaga2019}%
  \BibitemOpen
  \bibfield  {author} {\bibinfo {author} {\bibfnamefont {Y.}~\bibnamefont
  {Tokunaga}}, \bibinfo {author} {\bibfnamefont {H.}~\bibnamefont {Sakai}},
  \bibinfo {author} {\bibfnamefont {S.}~\bibnamefont {Kambe}}, \bibinfo
  {author} {\bibfnamefont {T.}~\bibnamefont {Hattori}}, \bibinfo {author}
  {\bibfnamefont {N.}~\bibnamefont {Higa}}, \bibinfo {author} {\bibfnamefont
  {G.}~\bibnamefont {Nakamine}}, \bibinfo {author} {\bibfnamefont
  {S.}~\bibnamefont {Kitagawa}}, \bibinfo {author} {\bibfnamefont
  {K.}~\bibnamefont {Ishida}}, \bibinfo {author} {\bibfnamefont
  {A.}~\bibnamefont {Nakamura}}, \bibinfo {author} {\bibfnamefont
  {Y.}~\bibnamefont {Shimizu}}, \bibinfo {author} {\bibfnamefont
  {Y.}~\bibnamefont {Homma}}, \bibinfo {author} {\bibfnamefont {D.~X.}\
  \bibnamefont {Li}}, \bibinfo {author} {\bibfnamefont {F.}~\bibnamefont
  {Honda}}, \ and\ \bibinfo {author} {\bibfnamefont {D.}~\bibnamefont {Aoki}},\
  }\href {\doibase 10.7566/JPSJ.88.073701} {\bibfield  {journal} {\bibinfo
  {journal} {J. Phys. Soc. Jpn.}\ }\textbf {\bibinfo {volume} {88}},\ \bibinfo
  {pages} {073701} (\bibinfo {year} {2019})}\BibitemShut {NoStop}%
\bibitem [{\citenamefont {Duan}\ \emph {et~al.}(2020)\citenamefont {Duan},
  \citenamefont {Sasmal}, \citenamefont {Maple}, \citenamefont {Podlesnyak},
  \citenamefont {Zhu}, \citenamefont {Si},\ and\ \citenamefont
  {Dai}}]{duan202incommensurate}%
  \BibitemOpen
  \bibfield  {author} {\bibinfo {author} {\bibfnamefont {C.}~\bibnamefont
  {Duan}}, \bibinfo {author} {\bibfnamefont {K.}~\bibnamefont {Sasmal}},
  \bibinfo {author} {\bibfnamefont {M.~B.}\ \bibnamefont {Maple}}, \bibinfo
  {author} {\bibfnamefont {A.}~\bibnamefont {Podlesnyak}}, \bibinfo {author}
  {\bibfnamefont {J.-X.}\ \bibnamefont {Zhu}}, \bibinfo {author} {\bibfnamefont
  {Q.}~\bibnamefont {Si}}, \ and\ \bibinfo {author} {\bibfnamefont
  {P.}~\bibnamefont {Dai}},\ }\href {\doibase 10.1103/PhysRevLett.125.237003}
  {\bibfield  {journal} {\bibinfo  {journal} {Phys. Rev. Lett.}\ }\textbf
  {\bibinfo {volume} {125}},\ \bibinfo {pages} {237003} (\bibinfo {year}
  {2020})}\BibitemShut {NoStop}%
\bibitem [{\citenamefont {Knafo}\ \emph {et~al.}(2021)\citenamefont {Knafo},
  \citenamefont {Knebel}, \citenamefont {Steffens}, \citenamefont {Kaneko},
  \citenamefont {Rosuel}, \citenamefont {Brison}, \citenamefont {Flouquet},
  \citenamefont {Aoki}, \citenamefont {Lapertot},\ and\ \citenamefont
  {Raymond}}]{Knafo2021}%
  \BibitemOpen
  \bibfield  {author} {\bibinfo {author} {\bibfnamefont {W.}~\bibnamefont
  {Knafo}}, \bibinfo {author} {\bibfnamefont {G.}~\bibnamefont {Knebel}},
  \bibinfo {author} {\bibfnamefont {P.}~\bibnamefont {Steffens}}, \bibinfo
  {author} {\bibfnamefont {K.}~\bibnamefont {Kaneko}}, \bibinfo {author}
  {\bibfnamefont {A.}~\bibnamefont {Rosuel}}, \bibinfo {author} {\bibfnamefont
  {J.-P.}\ \bibnamefont {Brison}}, \bibinfo {author} {\bibfnamefont
  {J.}~\bibnamefont {Flouquet}}, \bibinfo {author} {\bibfnamefont
  {D.}~\bibnamefont {Aoki}}, \bibinfo {author} {\bibfnamefont {G.}~\bibnamefont
  {Lapertot}}, \ and\ \bibinfo {author} {\bibfnamefont {S.}~\bibnamefont
  {Raymond}},\ }\href {\doibase 10.1103/PhysRevB.104.L100409} {\bibfield
  {journal} {\bibinfo  {journal} {Phys. Rev. B}\ }\textbf {\bibinfo {volume}
  {104}},\ \bibinfo {pages} {L100409} (\bibinfo {year} {2021})}\BibitemShut
  {NoStop}%
\bibitem [{\citenamefont {Duan}\ \emph {et~al.}(2021)\citenamefont {Duan},
  \citenamefont {Baumbach}, \citenamefont {Podlesnyak}, \citenamefont {Deng},
  \citenamefont {Moir}, \citenamefont {Breindel}, \citenamefont {Maple},
  \citenamefont {Nica}, \citenamefont {Si},\ and\ \citenamefont
  {Dai}}]{duan2021resonance}%
  \BibitemOpen
  \bibfield  {author} {\bibinfo {author} {\bibfnamefont {C.}~\bibnamefont
  {Duan}}, \bibinfo {author} {\bibfnamefont {R.~E.}\ \bibnamefont {Baumbach}},
  \bibinfo {author} {\bibfnamefont {A.}~\bibnamefont {Podlesnyak}}, \bibinfo
  {author} {\bibfnamefont {Y.}~\bibnamefont {Deng}}, \bibinfo {author}
  {\bibfnamefont {C.}~\bibnamefont {Moir}}, \bibinfo {author} {\bibfnamefont
  {A.~J.}\ \bibnamefont {Breindel}}, \bibinfo {author} {\bibfnamefont {M.~B.}\
  \bibnamefont {Maple}}, \bibinfo {author} {\bibfnamefont {E.~M.}\ \bibnamefont
  {Nica}}, \bibinfo {author} {\bibfnamefont {Q.}~\bibnamefont {Si}}, \ and\
  \bibinfo {author} {\bibfnamefont {P.}~\bibnamefont {Dai}},\ }\href
  {https://europepmc.org/article/med/34937893} {\bibfield  {journal} {\bibinfo
  {journal} {Nature}\ }\textbf {\bibinfo {volume} {600}},\ \bibinfo {pages}
  {636} (\bibinfo {year} {2021})}\BibitemShut {NoStop}%
\bibitem [{\citenamefont {Raymond}\ \emph {et~al.}(2021)\citenamefont
  {Raymond}, \citenamefont {Knafo}, \citenamefont {Knebel}, \citenamefont
  {Kaneko}, \citenamefont {Brison}, \citenamefont {Flouquet}, \citenamefont
  {Aoki},\ and\ \citenamefont {Lapertot}}]{raymond2021feedback}%
  \BibitemOpen
  \bibfield  {author} {\bibinfo {author} {\bibfnamefont {S.}~\bibnamefont
  {Raymond}}, \bibinfo {author} {\bibfnamefont {W.}~\bibnamefont {Knafo}},
  \bibinfo {author} {\bibfnamefont {G.}~\bibnamefont {Knebel}}, \bibinfo
  {author} {\bibfnamefont {K.}~\bibnamefont {Kaneko}}, \bibinfo {author}
  {\bibfnamefont {J.-P.}\ \bibnamefont {Brison}}, \bibinfo {author}
  {\bibfnamefont {J.}~\bibnamefont {Flouquet}}, \bibinfo {author}
  {\bibfnamefont {D.}~\bibnamefont {Aoki}}, \ and\ \bibinfo {author}
  {\bibfnamefont {G.}~\bibnamefont {Lapertot}},\ }\href
  {https://journals.jps.jp/doi/full/10.7566/JPSJ.90.113706} {\bibfield
  {journal} {\bibinfo  {journal} {J. Phys. Soc. Jpn.}\ }\textbf {\bibinfo
  {volume} {90}},\ \bibinfo {pages} {113706} (\bibinfo {year}
  {2021})}\BibitemShut {NoStop}%
\bibitem [{\citenamefont {Kreisel}\ \emph {et~al.}(2022)\citenamefont
  {Kreisel}, \citenamefont {Quan},\ and\ \citenamefont
  {Hirschfeld}}]{Kreisel2022}%
  \BibitemOpen
  \bibfield  {author} {\bibinfo {author} {\bibfnamefont {A.}~\bibnamefont
  {Kreisel}}, \bibinfo {author} {\bibfnamefont {Y.}~\bibnamefont {Quan}}, \
  and\ \bibinfo {author} {\bibfnamefont {P.~J.}\ \bibnamefont {Hirschfeld}},\
  }\href {\doibase 10.1103/PhysRevB.105.104507} {\bibfield  {journal} {\bibinfo
   {journal} {Phys. Rev. B}\ }\textbf {\bibinfo {volume} {105}},\ \bibinfo
  {pages} {104507} (\bibinfo {year} {2022})}\BibitemShut {NoStop}%
\bibitem [{\citenamefont {Miao}\ \emph {et~al.}(2020)\citenamefont {Miao},
  \citenamefont {Liu}, \citenamefont {Xu}, \citenamefont {Kotta}, \citenamefont
  {Kang}, \citenamefont {Ran}, \citenamefont {Paglione}, \citenamefont
  {Kotliar}, \citenamefont {Butch}, \citenamefont {Denlinger},\ and\
  \citenamefont {Wray}}]{Miao2020ARPES}%
  \BibitemOpen
  \bibfield  {author} {\bibinfo {author} {\bibfnamefont {L.}~\bibnamefont
  {Miao}}, \bibinfo {author} {\bibfnamefont {S.}~\bibnamefont {Liu}}, \bibinfo
  {author} {\bibfnamefont {Y.}~\bibnamefont {Xu}}, \bibinfo {author}
  {\bibfnamefont {E.~C.}\ \bibnamefont {Kotta}}, \bibinfo {author}
  {\bibfnamefont {C.-J.}\ \bibnamefont {Kang}}, \bibinfo {author}
  {\bibfnamefont {S.}~\bibnamefont {Ran}}, \bibinfo {author} {\bibfnamefont
  {J.}~\bibnamefont {Paglione}}, \bibinfo {author} {\bibfnamefont
  {G.}~\bibnamefont {Kotliar}}, \bibinfo {author} {\bibfnamefont {N.~P.}\
  \bibnamefont {Butch}}, \bibinfo {author} {\bibfnamefont {J.~D.}\ \bibnamefont
  {Denlinger}}, \ and\ \bibinfo {author} {\bibfnamefont {L.~A.}\ \bibnamefont
  {Wray}},\ }\href {\doibase 10.1103/PhysRevLett.124.076401} {\bibfield
  {journal} {\bibinfo  {journal} {Phys. Rev. Lett.}\ }\textbf {\bibinfo
  {volume} {124}},\ \bibinfo {pages} {076401} (\bibinfo {year}
  {2020})}\BibitemShut {NoStop}%
\bibitem [{\citenamefont {Xu}\ \emph {et~al.}(2019)\citenamefont {Xu},
  \citenamefont {Sheng},\ and\ \citenamefont {Yang}}]{Xu2019DMFT}%
  \BibitemOpen
  \bibfield  {author} {\bibinfo {author} {\bibfnamefont {Y.}~\bibnamefont
  {Xu}}, \bibinfo {author} {\bibfnamefont {Y.}~\bibnamefont {Sheng}}, \ and\
  \bibinfo {author} {\bibfnamefont {Y.-f.}\ \bibnamefont {Yang}},\ }\href
  {\doibase 10.1103/PhysRevLett.123.217002} {\bibfield  {journal} {\bibinfo
  {journal} {Phys. Rev. Lett.}\ }\textbf {\bibinfo {volume} {123}},\ \bibinfo
  {pages} {217002} (\bibinfo {year} {2019})}\BibitemShut {NoStop}%
\bibitem [{\citenamefont {Fujimori}\ \emph {et~al.}(2021)\citenamefont
  {Fujimori}, \citenamefont {Kawasaki}, \citenamefont {Takeda}, \citenamefont
  {Yamagami}, \citenamefont {Nakamura}, \citenamefont {Homma},\ and\
  \citenamefont {Aoki}}]{fujimori2021}%
  \BibitemOpen
  \bibfield  {author} {\bibinfo {author} {\bibfnamefont {S.-i.}\ \bibnamefont
  {Fujimori}}, \bibinfo {author} {\bibfnamefont {I.}~\bibnamefont {Kawasaki}},
  \bibinfo {author} {\bibfnamefont {Y.}~\bibnamefont {Takeda}}, \bibinfo
  {author} {\bibfnamefont {H.}~\bibnamefont {Yamagami}}, \bibinfo {author}
  {\bibfnamefont {A.}~\bibnamefont {Nakamura}}, \bibinfo {author}
  {\bibfnamefont {Y.}~\bibnamefont {Homma}}, \ and\ \bibinfo {author}
  {\bibfnamefont {D.}~\bibnamefont {Aoki}},\ }\href {\doibase
  10.7566/JPSJ.90.015002} {\bibfield  {journal} {\bibinfo  {journal} {Journal
  of the Physical Society of Japan}\ }\textbf {\bibinfo {volume} {90}},\
  \bibinfo {pages} {015002} (\bibinfo {year} {2021})},\ \Eprint
  {http://arxiv.org/abs/https://doi.org/10.7566/JPSJ.90.015002}
  {https://doi.org/10.7566/JPSJ.90.015002} \BibitemShut {NoStop}%
\bibitem [{\citenamefont {Sakai}\ \emph {et~al.}(2022)\citenamefont {Sakai},
  \citenamefont {Opletal}, \citenamefont {Tokiwa}, \citenamefont {Yamamoto},
  \citenamefont {Tokunaga}, \citenamefont {Kambe},\ and\ \citenamefont
  {Haga}}]{Sakai2022crystal}%
  \BibitemOpen
  \bibfield  {author} {\bibinfo {author} {\bibfnamefont {H.}~\bibnamefont
  {Sakai}}, \bibinfo {author} {\bibfnamefont {P.}~\bibnamefont {Opletal}},
  \bibinfo {author} {\bibfnamefont {Y.}~\bibnamefont {Tokiwa}}, \bibinfo
  {author} {\bibfnamefont {E.}~\bibnamefont {Yamamoto}}, \bibinfo {author}
  {\bibfnamefont {Y.}~\bibnamefont {Tokunaga}}, \bibinfo {author}
  {\bibfnamefont {S.}~\bibnamefont {Kambe}}, \ and\ \bibinfo {author}
  {\bibfnamefont {Y.}~\bibnamefont {Haga}},\ }\href {\doibase
  10.1103/PhysRevMaterials.6.073401} {\bibfield  {journal} {\bibinfo  {journal}
  {Phys. Rev. Mater.}\ }\textbf {\bibinfo {volume} {6}},\ \bibinfo {pages}
  {073401} (\bibinfo {year} {2022})}\BibitemShut {NoStop}%
\bibitem [{\citenamefont {Aoki}\ \emph
  {et~al.}(2022{\natexlab{b}})\citenamefont {Aoki}, \citenamefont {Sakai},
  \citenamefont {Opletal}, \citenamefont {Tokiwa}, \citenamefont {Ishizuka},
  \citenamefont {Yanase}, \citenamefont {Harima}, \citenamefont {Nakamura},
  \citenamefont {Li}, \citenamefont {Homma}, \citenamefont {Shimizu},
  \citenamefont {Knebel}, \citenamefont {Flouquet},\ and\ \citenamefont
  {Haga}}]{Aoki2022dHvA}%
  \BibitemOpen
  \bibfield  {author} {\bibinfo {author} {\bibfnamefont {D.}~\bibnamefont
  {Aoki}}, \bibinfo {author} {\bibfnamefont {H.}~\bibnamefont {Sakai}},
  \bibinfo {author} {\bibfnamefont {P.}~\bibnamefont {Opletal}}, \bibinfo
  {author} {\bibfnamefont {Y.}~\bibnamefont {Tokiwa}}, \bibinfo {author}
  {\bibfnamefont {J.}~\bibnamefont {Ishizuka}}, \bibinfo {author}
  {\bibfnamefont {Y.}~\bibnamefont {Yanase}}, \bibinfo {author} {\bibfnamefont
  {H.}~\bibnamefont {Harima}}, \bibinfo {author} {\bibfnamefont
  {A.}~\bibnamefont {Nakamura}}, \bibinfo {author} {\bibfnamefont
  {D.}~\bibnamefont {Li}}, \bibinfo {author} {\bibfnamefont {Y.}~\bibnamefont
  {Homma}}, \bibinfo {author} {\bibfnamefont {Y.}~\bibnamefont {Shimizu}},
  \bibinfo {author} {\bibfnamefont {G.}~\bibnamefont {Knebel}}, \bibinfo
  {author} {\bibfnamefont {J.}~\bibnamefont {Flouquet}}, \ and\ \bibinfo
  {author} {\bibfnamefont {Y.}~\bibnamefont {Haga}},\ }\href {\doibase
  10.7566/JPSJ.91.083704} {\bibfield  {journal} {\bibinfo  {journal} {Journal
  of the Physical Society of Japan}\ }\textbf {\bibinfo {volume} {91}},\
  \bibinfo {pages} {083704} (\bibinfo {year} {2022}{\natexlab{b}})},\ \Eprint
  {http://arxiv.org/abs/https://doi.org/10.7566/JPSJ.91.083704}
  {https://doi.org/10.7566/JPSJ.91.083704} \BibitemShut {NoStop}%
\bibitem [{\citenamefont {Eaton}\ \emph {et~al.}(2023)\citenamefont {Eaton},
  \citenamefont {Weinberger}, \citenamefont {Popiel}, \citenamefont {Wu},
  \citenamefont {Hickey}, \citenamefont {Cabala}, \citenamefont {Pospisil},
  \citenamefont {Prokleska}, \citenamefont {Haidamak}, \citenamefont {Bastien},
  \citenamefont {Opletal}, \citenamefont {Sakai}, \citenamefont {Haga},
  \citenamefont {Nowell}, \citenamefont {Benjamin}, \citenamefont {Sechovsky},
  \citenamefont {Lonzarich}, \citenamefont {Grosche},\ and\ \citenamefont
  {Valiska}}]{eaton2023quasi2d}%
  \BibitemOpen
  \bibfield  {author} {\bibinfo {author} {\bibfnamefont {A.~G.}\ \bibnamefont
  {Eaton}}, \bibinfo {author} {\bibfnamefont {T.~I.}\ \bibnamefont
  {Weinberger}}, \bibinfo {author} {\bibfnamefont {N.~J.~M.}\ \bibnamefont
  {Popiel}}, \bibinfo {author} {\bibfnamefont {Z.}~\bibnamefont {Wu}}, \bibinfo
  {author} {\bibfnamefont {A.~J.}\ \bibnamefont {Hickey}}, \bibinfo {author}
  {\bibfnamefont {A.}~\bibnamefont {Cabala}}, \bibinfo {author} {\bibfnamefont
  {J.}~\bibnamefont {Pospisil}}, \bibinfo {author} {\bibfnamefont
  {J.}~\bibnamefont {Prokleska}}, \bibinfo {author} {\bibfnamefont
  {T.}~\bibnamefont {Haidamak}}, \bibinfo {author} {\bibfnamefont
  {G.}~\bibnamefont {Bastien}}, \bibinfo {author} {\bibfnamefont
  {P.}~\bibnamefont {Opletal}}, \bibinfo {author} {\bibfnamefont
  {H.}~\bibnamefont {Sakai}}, \bibinfo {author} {\bibfnamefont
  {Y.}~\bibnamefont {Haga}}, \bibinfo {author} {\bibfnamefont {R.}~\bibnamefont
  {Nowell}}, \bibinfo {author} {\bibfnamefont {S.~M.}\ \bibnamefont
  {Benjamin}}, \bibinfo {author} {\bibfnamefont {V.}~\bibnamefont {Sechovsky}},
  \bibinfo {author} {\bibfnamefont {G.~G.}\ \bibnamefont {Lonzarich}}, \bibinfo
  {author} {\bibfnamefont {F.~M.}\ \bibnamefont {Grosche}}, \ and\ \bibinfo
  {author} {\bibfnamefont {M.}~\bibnamefont {Valiska}},\ }\href@noop {}
  {\enquote {\bibinfo {title} {Quasi-2d fermi surface in the anomalous
  superconductor ute$_2$},}\ } (\bibinfo {year} {2023}),\ \Eprint
  {http://arxiv.org/abs/2302.04758} {arXiv:2302.04758 [cond-mat.supr-con]}
  \BibitemShut {NoStop}%
\bibitem [{\citenamefont {Broyles}\ \emph {et~al.}(2023)\citenamefont
  {Broyles}, \citenamefont {Rehfuss}, \citenamefont {Siddiquee}, \citenamefont
  {Zheng}, \citenamefont {Le}, \citenamefont {Nikolo}, \citenamefont {Graf},
  \citenamefont {Singleton},\ and\ \citenamefont {Ran}}]{broyles2023revealing}%
  \BibitemOpen
  \bibfield  {author} {\bibinfo {author} {\bibfnamefont {C.}~\bibnamefont
  {Broyles}}, \bibinfo {author} {\bibfnamefont {Z.}~\bibnamefont {Rehfuss}},
  \bibinfo {author} {\bibfnamefont {H.}~\bibnamefont {Siddiquee}}, \bibinfo
  {author} {\bibfnamefont {K.}~\bibnamefont {Zheng}}, \bibinfo {author}
  {\bibfnamefont {Y.}~\bibnamefont {Le}}, \bibinfo {author} {\bibfnamefont
  {M.}~\bibnamefont {Nikolo}}, \bibinfo {author} {\bibfnamefont
  {D.}~\bibnamefont {Graf}}, \bibinfo {author} {\bibfnamefont {J.}~\bibnamefont
  {Singleton}}, \ and\ \bibinfo {author} {\bibfnamefont {S.}~\bibnamefont
  {Ran}},\ }\href@noop {} {\enquote {\bibinfo {title} {Revealing a 3d fermi
  surface and electron-hole tunneling in ute$_{2}$ with quantum
  oscillations},}\ } (\bibinfo {year} {2023}),\ \Eprint
  {http://arxiv.org/abs/2303.09050} {arXiv:2303.09050 [cond-mat.str-el]}
  \BibitemShut {NoStop}%
\bibitem [{\citenamefont {Aoki}\ \emph {et~al.}(2023)\citenamefont {Aoki},
  \citenamefont {Sheikin}, \citenamefont {McCollam}, \citenamefont {Ishizuka},
  \citenamefont {Yanase}, \citenamefont {Lapertot}, \citenamefont {Flouquet},\
  and\ \citenamefont {Knebel}}]{aoki2023haasvan}%
  \BibitemOpen
  \bibfield  {author} {\bibinfo {author} {\bibfnamefont {D.}~\bibnamefont
  {Aoki}}, \bibinfo {author} {\bibfnamefont {I.}~\bibnamefont {Sheikin}},
  \bibinfo {author} {\bibfnamefont {A.}~\bibnamefont {McCollam}}, \bibinfo
  {author} {\bibfnamefont {J.}~\bibnamefont {Ishizuka}}, \bibinfo {author}
  {\bibfnamefont {Y.}~\bibnamefont {Yanase}}, \bibinfo {author} {\bibfnamefont
  {G.}~\bibnamefont {Lapertot}}, \bibinfo {author} {\bibfnamefont
  {J.}~\bibnamefont {Flouquet}}, \ and\ \bibinfo {author} {\bibfnamefont
  {G.}~\bibnamefont {Knebel}},\ }\href {\doibase 10.7566/JPSJ.92.065002}
  {\bibfield  {journal} {\bibinfo  {journal} {Journal of the Physical Society
  of Japan}\ }\textbf {\bibinfo {volume} {92}},\ \bibinfo {pages} {065002}
  (\bibinfo {year} {2023})},\ \Eprint
  {http://arxiv.org/abs/https://doi.org/10.7566/JPSJ.92.065002}
  {https://doi.org/10.7566/JPSJ.92.065002} \BibitemShut {NoStop}%
\bibitem [{\citenamefont {Aoki}\ \emph {et~al.}(2019)\citenamefont {Aoki},
  \citenamefont {Ishida},\ and\ \citenamefont {Flouquet}}]{aokiFMSCreview}%
  \BibitemOpen
  \bibfield  {author} {\bibinfo {author} {\bibfnamefont {D.}~\bibnamefont
  {Aoki}}, \bibinfo {author} {\bibfnamefont {K.}~\bibnamefont {Ishida}}, \ and\
  \bibinfo {author} {\bibfnamefont {J.}~\bibnamefont {Flouquet}},\ }\href
  {\doibase 10.7566/JPSJ.88.022001} {\bibfield  {journal} {\bibinfo  {journal}
  {Journal of the Physical Society of Japan}\ }\textbf {\bibinfo {volume}
  {88}},\ \bibinfo {pages} {022001} (\bibinfo {year} {2019})},\ \Eprint
  {http://arxiv.org/abs/https://doi.org/10.7566/JPSJ.88.022001}
  {https://doi.org/10.7566/JPSJ.88.022001} \BibitemShut {NoStop}%
\bibitem [{\citenamefont {Braithwaite}\ \emph {et~al.}(2019)\citenamefont
  {Braithwaite}, \citenamefont {Vali{\v{s}}ka}, \citenamefont {Knebel},
  \citenamefont {Lapertot}, \citenamefont {Brison}, \citenamefont {Pourret},
  \citenamefont {Zhitomirsky}, \citenamefont {Flouquet}, \citenamefont
  {Honda},\ and\ \citenamefont {Aoki}}]{Braithwaite2019}%
  \BibitemOpen
  \bibfield  {author} {\bibinfo {author} {\bibfnamefont {D.}~\bibnamefont
  {Braithwaite}}, \bibinfo {author} {\bibfnamefont {M.}~\bibnamefont
  {Vali{\v{s}}ka}}, \bibinfo {author} {\bibfnamefont {G.}~\bibnamefont
  {Knebel}}, \bibinfo {author} {\bibfnamefont {G.}~\bibnamefont {Lapertot}},
  \bibinfo {author} {\bibfnamefont {J.~P.}\ \bibnamefont {Brison}}, \bibinfo
  {author} {\bibfnamefont {A.}~\bibnamefont {Pourret}}, \bibinfo {author}
  {\bibfnamefont {M.~E.}\ \bibnamefont {Zhitomirsky}}, \bibinfo {author}
  {\bibfnamefont {J.}~\bibnamefont {Flouquet}}, \bibinfo {author}
  {\bibfnamefont {F.}~\bibnamefont {Honda}}, \ and\ \bibinfo {author}
  {\bibfnamefont {D.}~\bibnamefont {Aoki}},\ }\href {\doibase
  10.1038/s42005-019-0248-z} {\bibfield  {journal} {\bibinfo  {journal}
  {Communications Physics}\ }\textbf {\bibinfo {volume} {2}},\ \bibinfo {pages}
  {1} (\bibinfo {year} {2019})}\BibitemShut {NoStop}%
\bibitem [{\citenamefont {Aoki}\ \emph {et~al.}(2020)\citenamefont {Aoki},
  \citenamefont {Honda}, \citenamefont {Knebel}, \citenamefont {Braithwaite},
  \citenamefont {Nakamura}, \citenamefont {Li}, \citenamefont {Homma},
  \citenamefont {Shimizu}, \citenamefont {Sato}, \citenamefont {Brison},\ and\
  \citenamefont {Flouquet}}]{Aoki2020}%
  \BibitemOpen
  \bibfield  {author} {\bibinfo {author} {\bibfnamefont {D.}~\bibnamefont
  {Aoki}}, \bibinfo {author} {\bibfnamefont {F.}~\bibnamefont {Honda}},
  \bibinfo {author} {\bibfnamefont {G.}~\bibnamefont {Knebel}}, \bibinfo
  {author} {\bibfnamefont {D.}~\bibnamefont {Braithwaite}}, \bibinfo {author}
  {\bibfnamefont {A.}~\bibnamefont {Nakamura}}, \bibinfo {author}
  {\bibfnamefont {D.}~\bibnamefont {Li}}, \bibinfo {author} {\bibfnamefont
  {Y.}~\bibnamefont {Homma}}, \bibinfo {author} {\bibfnamefont
  {Y.}~\bibnamefont {Shimizu}}, \bibinfo {author} {\bibfnamefont {Y.~J.}\
  \bibnamefont {Sato}}, \bibinfo {author} {\bibfnamefont {J.-P.}\ \bibnamefont
  {Brison}}, \ and\ \bibinfo {author} {\bibfnamefont {J.}~\bibnamefont
  {Flouquet}},\ }\href {\doibase 10.7566/JPSJ.89.053705} {\bibfield  {journal}
  {\bibinfo  {journal} {J. Phys. Soc. Jpn.}\ }\textbf {\bibinfo {volume}
  {89}},\ \bibinfo {pages} {053705} (\bibinfo {year} {2020})}\BibitemShut
  {NoStop}%
\bibitem [{\citenamefont {Lin}\ \emph {et~al.}(2020)\citenamefont {Lin},
  \citenamefont {Campbell}, \citenamefont {Ran}, \citenamefont {Liu},
  \citenamefont {Kim}, \citenamefont {Nevidomskyy}, \citenamefont {Graf},
  \citenamefont {Butch},\ and\ \citenamefont {Paglione}}]{Lin2020}%
  \BibitemOpen
  \bibfield  {author} {\bibinfo {author} {\bibfnamefont {W.~C.}\ \bibnamefont
  {Lin}}, \bibinfo {author} {\bibfnamefont {D.~J.}\ \bibnamefont {Campbell}},
  \bibinfo {author} {\bibfnamefont {S.}~\bibnamefont {Ran}}, \bibinfo {author}
  {\bibfnamefont {I.~L.}\ \bibnamefont {Liu}}, \bibinfo {author} {\bibfnamefont
  {H.}~\bibnamefont {Kim}}, \bibinfo {author} {\bibfnamefont {A.~H.}\
  \bibnamefont {Nevidomskyy}}, \bibinfo {author} {\bibfnamefont
  {D.}~\bibnamefont {Graf}}, \bibinfo {author} {\bibfnamefont {N.~P.}\
  \bibnamefont {Butch}}, \ and\ \bibinfo {author} {\bibfnamefont
  {J.}~\bibnamefont {Paglione}},\ }\href {\doibase 10.1038/s41535-020-00270-w}
  {\bibfield  {journal} {\bibinfo  {journal} {npj Quantum Materials}\ }\textbf
  {\bibinfo {volume} {5}},\ \bibinfo {pages} {1} (\bibinfo {year}
  {2020})}\BibitemShut {NoStop}%
\bibitem [{\citenamefont {Ran}\ \emph {et~al.}(2020)\citenamefont {Ran},
  \citenamefont {Kim}, \citenamefont {Liu}, \citenamefont {Saha}, \citenamefont
  {Hayes}, \citenamefont {Metz}, \citenamefont {Eo}, \citenamefont {Paglione},\
  and\ \citenamefont {Butch}}]{RanPRB2020}%
  \BibitemOpen
  \bibfield  {author} {\bibinfo {author} {\bibfnamefont {S.}~\bibnamefont
  {Ran}}, \bibinfo {author} {\bibfnamefont {H.}~\bibnamefont {Kim}}, \bibinfo
  {author} {\bibfnamefont {I.~L.}\ \bibnamefont {Liu}}, \bibinfo {author}
  {\bibfnamefont {S.~R.}\ \bibnamefont {Saha}}, \bibinfo {author}
  {\bibfnamefont {I.}~\bibnamefont {Hayes}}, \bibinfo {author} {\bibfnamefont
  {T.}~\bibnamefont {Metz}}, \bibinfo {author} {\bibfnamefont {Y.~S.}\
  \bibnamefont {Eo}}, \bibinfo {author} {\bibfnamefont {J.}~\bibnamefont
  {Paglione}}, \ and\ \bibinfo {author} {\bibfnamefont {N.~P.}\ \bibnamefont
  {Butch}},\ }\href {\doibase 10.1103/PhysRevB.101.140503} {\bibfield
  {journal} {\bibinfo  {journal} {Phys. Rev. B}\ }\textbf {\bibinfo {volume}
  {101}},\ \bibinfo {pages} {140503(R)} (\bibinfo {year} {2020})}\BibitemShut
  {NoStop}%
\bibitem [{\citenamefont {Thomas}\ \emph {et~al.}(2020)\citenamefont {Thomas},
  \citenamefont {Santos}, \citenamefont {Christensen}, \citenamefont {Asaba},
  \citenamefont {Ronning}, \citenamefont {Thompson}, \citenamefont {Bauer},
  \citenamefont {Fernandes}, \citenamefont {Fabbris},\ and\ \citenamefont
  {Rosa}}]{Thomas2020}%
  \BibitemOpen
  \bibfield  {author} {\bibinfo {author} {\bibfnamefont {S.~M.}\ \bibnamefont
  {Thomas}}, \bibinfo {author} {\bibfnamefont {F.~B.}\ \bibnamefont {Santos}},
  \bibinfo {author} {\bibfnamefont {M.~H.}\ \bibnamefont {Christensen}},
  \bibinfo {author} {\bibfnamefont {T.}~\bibnamefont {Asaba}}, \bibinfo
  {author} {\bibfnamefont {F.}~\bibnamefont {Ronning}}, \bibinfo {author}
  {\bibfnamefont {J.~D.}\ \bibnamefont {Thompson}}, \bibinfo {author}
  {\bibfnamefont {E.~D.}\ \bibnamefont {Bauer}}, \bibinfo {author}
  {\bibfnamefont {R.~M.}\ \bibnamefont {Fernandes}}, \bibinfo {author}
  {\bibfnamefont {G.}~\bibnamefont {Fabbris}}, \ and\ \bibinfo {author}
  {\bibfnamefont {P.~F.}\ \bibnamefont {Rosa}},\ }\href {\doibase
  10.1126/sciadv.abc8709} {\bibfield  {journal} {\bibinfo  {journal} {Sci.
  Adv.}\ }\textbf {\bibinfo {volume} {6}},\ \bibinfo {pages} {eabc8709}
  (\bibinfo {year} {2020})}\BibitemShut {NoStop}%
\bibitem [{\citenamefont {Knebel}\ \emph {et~al.}(2020)\citenamefont {Knebel},
  \citenamefont {Kimata}, \citenamefont {Vali{\v{s}}ka}, \citenamefont {Honda},
  \citenamefont {Li}, \citenamefont {Braithwaite}, \citenamefont {Lapertot},
  \citenamefont {Knafo}, \citenamefont {Pourret}, \citenamefont {Sato},
  \citenamefont {Shimizu}, \citenamefont {Kihara}, \citenamefont {Brison},
  \citenamefont {Flouquet},\ and\ \citenamefont {Aoki}}]{Knebel2020}%
  \BibitemOpen
  \bibfield  {author} {\bibinfo {author} {\bibfnamefont {G.}~\bibnamefont
  {Knebel}}, \bibinfo {author} {\bibfnamefont {M.}~\bibnamefont {Kimata}},
  \bibinfo {author} {\bibfnamefont {M.}~\bibnamefont {Vali{\v{s}}ka}}, \bibinfo
  {author} {\bibfnamefont {F.}~\bibnamefont {Honda}}, \bibinfo {author}
  {\bibfnamefont {D.}~\bibnamefont {Li}}, \bibinfo {author} {\bibfnamefont
  {D.}~\bibnamefont {Braithwaite}}, \bibinfo {author} {\bibfnamefont
  {G.}~\bibnamefont {Lapertot}}, \bibinfo {author} {\bibfnamefont
  {W.}~\bibnamefont {Knafo}}, \bibinfo {author} {\bibfnamefont
  {A.}~\bibnamefont {Pourret}}, \bibinfo {author} {\bibfnamefont {Y.~J.}\
  \bibnamefont {Sato}}, \bibinfo {author} {\bibfnamefont {Y.}~\bibnamefont
  {Shimizu}}, \bibinfo {author} {\bibfnamefont {T.}~\bibnamefont {Kihara}},
  \bibinfo {author} {\bibfnamefont {J.-P.}\ \bibnamefont {Brison}}, \bibinfo
  {author} {\bibfnamefont {J.}~\bibnamefont {Flouquet}}, \ and\ \bibinfo
  {author} {\bibfnamefont {D.}~\bibnamefont {Aoki}},\ }\href {\doibase
  10.7566/JPSJ.89.053707} {\bibfield  {journal} {\bibinfo  {journal} {J. Phys.
  Soc. Jpn.}\ }\textbf {\bibinfo {volume} {89}},\ \bibinfo {pages} {053707}
  (\bibinfo {year} {2020})}\BibitemShut {NoStop}%
\bibitem [{\citenamefont {Aoki}\ \emph {et~al.}(2021)\citenamefont {Aoki},
  \citenamefont {Kimata}, \citenamefont {Sato}, \citenamefont {Knebel},
  \citenamefont {Honda}, \citenamefont {Nakamura}, \citenamefont {Li},
  \citenamefont {Homma}, \citenamefont {Shimizu}, \citenamefont {Knafo},
  \citenamefont {Braithwaite}, \citenamefont {Vališka}, \citenamefont
  {Pourret}, \citenamefont {Brison},\ and\ \citenamefont
  {Flouquet}}]{aoki2021fieldinduced}%
  \BibitemOpen
  \bibfield  {author} {\bibinfo {author} {\bibfnamefont {D.}~\bibnamefont
  {Aoki}}, \bibinfo {author} {\bibfnamefont {M.}~\bibnamefont {Kimata}},
  \bibinfo {author} {\bibfnamefont {Y.~J.}\ \bibnamefont {Sato}}, \bibinfo
  {author} {\bibfnamefont {G.}~\bibnamefont {Knebel}}, \bibinfo {author}
  {\bibfnamefont {F.}~\bibnamefont {Honda}}, \bibinfo {author} {\bibfnamefont
  {A.}~\bibnamefont {Nakamura}}, \bibinfo {author} {\bibfnamefont
  {D.}~\bibnamefont {Li}}, \bibinfo {author} {\bibfnamefont {Y.}~\bibnamefont
  {Homma}}, \bibinfo {author} {\bibfnamefont {Y.}~\bibnamefont {Shimizu}},
  \bibinfo {author} {\bibfnamefont {W.}~\bibnamefont {Knafo}}, \bibinfo
  {author} {\bibfnamefont {D.}~\bibnamefont {Braithwaite}}, \bibinfo {author}
  {\bibfnamefont {M.}~\bibnamefont {Vališka}}, \bibinfo {author}
  {\bibfnamefont {A.}~\bibnamefont {Pourret}}, \bibinfo {author} {\bibfnamefont
  {J.-P.}\ \bibnamefont {Brison}}, \ and\ \bibinfo {author} {\bibfnamefont
  {J.}~\bibnamefont {Flouquet}},\ }\href {\doibase 10.7566/JPSJ.90.074705}
  {\bibfield  {journal} {\bibinfo  {journal} {Journal of the Physical Society
  of Japan}\ }\textbf {\bibinfo {volume} {90}},\ \bibinfo {pages} {074705}
  (\bibinfo {year} {2021})},\ \Eprint
  {http://arxiv.org/abs/https://doi.org/10.7566/JPSJ.90.074705}
  {https://doi.org/10.7566/JPSJ.90.074705} \BibitemShut {NoStop}%
\bibitem [{\citenamefont {Rosa}\ \emph {et~al.}(2022)\citenamefont {Rosa},
  \citenamefont {Weiland}, \citenamefont {Fender}, \citenamefont {Scott},
  \citenamefont {Ronning}, \citenamefont {Thompson}, \citenamefont {Bauer},\
  and\ \citenamefont {Thomas}}]{Rosa2022}%
  \BibitemOpen
  \bibfield  {author} {\bibinfo {author} {\bibfnamefont {P.~F.~S.}\
  \bibnamefont {Rosa}}, \bibinfo {author} {\bibfnamefont {A.}~\bibnamefont
  {Weiland}}, \bibinfo {author} {\bibfnamefont {S.~S.}\ \bibnamefont {Fender}},
  \bibinfo {author} {\bibfnamefont {B.~L.}\ \bibnamefont {Scott}}, \bibinfo
  {author} {\bibfnamefont {F.}~\bibnamefont {Ronning}}, \bibinfo {author}
  {\bibfnamefont {J.~D.}\ \bibnamefont {Thompson}}, \bibinfo {author}
  {\bibfnamefont {E.~D.}\ \bibnamefont {Bauer}}, \ and\ \bibinfo {author}
  {\bibfnamefont {S.~M.}\ \bibnamefont {Thomas}},\ }\href {\doibase
  10.1038/s43246-022-00254-2} {\bibfield  {journal} {\bibinfo  {journal}
  {Communications Materials}\ }\textbf {\bibinfo {volume} {3}},\ \bibinfo
  {pages} {33} (\bibinfo {year} {2022})}\BibitemShut {NoStop}%
\bibitem [{\citenamefont {Ishizuka}\ and\ \citenamefont
  {Yanase}(2021)}]{Ishizuka2020}%
  \BibitemOpen
  \bibfield  {author} {\bibinfo {author} {\bibfnamefont {J.}~\bibnamefont
  {Ishizuka}}\ and\ \bibinfo {author} {\bibfnamefont {Y.}~\bibnamefont
  {Yanase}},\ }\href {\doibase 10.1103/PhysRevB.103.094504} {\bibfield
  {journal} {\bibinfo  {journal} {Phys. Rev. B}\ }\textbf {\bibinfo {volume}
  {103}},\ \bibinfo {pages} {094504} (\bibinfo {year} {2021})}\BibitemShut
  {NoStop}%
\bibitem [{\citenamefont {Shishidou}\ \emph {et~al.}(2021)\citenamefont
  {Shishidou}, \citenamefont {Suh}, \citenamefont {Brydon}, \citenamefont
  {Weinert},\ and\ \citenamefont {Agterberg}}]{Shishidou2021}%
  \BibitemOpen
  \bibfield  {author} {\bibinfo {author} {\bibfnamefont {T.}~\bibnamefont
  {Shishidou}}, \bibinfo {author} {\bibfnamefont {H.~G.}\ \bibnamefont {Suh}},
  \bibinfo {author} {\bibfnamefont {P.~M.~R.}\ \bibnamefont {Brydon}}, \bibinfo
  {author} {\bibfnamefont {M.}~\bibnamefont {Weinert}}, \ and\ \bibinfo
  {author} {\bibfnamefont {D.~F.}\ \bibnamefont {Agterberg}},\ }\href {\doibase
  10.1103/PhysRevB.103.104504} {\bibfield  {journal} {\bibinfo  {journal}
  {Phys. Rev. B}\ }\textbf {\bibinfo {volume} {103}},\ \bibinfo {pages}
  {104504} (\bibinfo {year} {2021})}\BibitemShut {NoStop}%
\bibitem [{\citenamefont {Hazra}\ and\ \citenamefont
  {Coleman}(2023)}]{Hazra2023}%
  \BibitemOpen
  \bibfield  {author} {\bibinfo {author} {\bibfnamefont {T.}~\bibnamefont
  {Hazra}}\ and\ \bibinfo {author} {\bibfnamefont {P.}~\bibnamefont
  {Coleman}},\ }\href {\doibase 10.1103/PhysRevLett.130.136002} {\bibfield
  {journal} {\bibinfo  {journal} {Phys. Rev. Lett.}\ }\textbf {\bibinfo
  {volume} {130}},\ \bibinfo {pages} {136002} (\bibinfo {year}
  {2023})}\BibitemShut {NoStop}%
\bibitem [{\citenamefont {Chen}\ \emph {et~al.}(2021)\citenamefont {Chen},
  \citenamefont {Hu}, \citenamefont {Lane}, \citenamefont {Nica}, \citenamefont
  {Zhu},\ and\ \citenamefont {Si}}]{chen2021multiorbital}%
  \BibitemOpen
  \bibfield  {author} {\bibinfo {author} {\bibfnamefont {L.}~\bibnamefont
  {Chen}}, \bibinfo {author} {\bibfnamefont {H.}~\bibnamefont {Hu}}, \bibinfo
  {author} {\bibfnamefont {C.}~\bibnamefont {Lane}}, \bibinfo {author}
  {\bibfnamefont {E.~M.}\ \bibnamefont {Nica}}, \bibinfo {author}
  {\bibfnamefont {J.-X.}\ \bibnamefont {Zhu}}, \ and\ \bibinfo {author}
  {\bibfnamefont {Q.}~\bibnamefont {Si}},\ }\href@noop {} {\enquote {\bibinfo
  {title} {Multiorbital spin-triplet pairing and spin resonance in the
  heavy-fermion superconductor $\mathrm{UTe_2}$},}\ } (\bibinfo {year}
  {2021}),\ \Eprint {http://arxiv.org/abs/2112.14750} {arXiv:2112.14750
  [cond-mat.supr-con]} \BibitemShut {NoStop}%
\bibitem [{\citenamefont {Choi}\ \emph {et~al.}(2023)\citenamefont {Choi},
  \citenamefont {Lee},\ and\ \citenamefont {Yang}}]{choi2023correlated}%
  \BibitemOpen
  \bibfield  {author} {\bibinfo {author} {\bibfnamefont {H.~C.}\ \bibnamefont
  {Choi}}, \bibinfo {author} {\bibfnamefont {S.~H.}\ \bibnamefont {Lee}}, \
  and\ \bibinfo {author} {\bibfnamefont {B.-J.}\ \bibnamefont {Yang}},\
  }\href@noop {} {\enquote {\bibinfo {title} {Correlated normal state
  fermiology and topological superconductivity in ute2},}\ } (\bibinfo {year}
  {2023}),\ \Eprint {http://arxiv.org/abs/2206.04876} {arXiv:2206.04876
  [cond-mat.supr-con]} \BibitemShut {NoStop}%
\bibitem [{\citenamefont {Yu}\ \emph {et~al.}(2023)\citenamefont {Yu},
  \citenamefont {Yu}, \citenamefont {Agterberg},\ and\ \citenamefont
  {Raghu}}]{yu2023theory}%
  \BibitemOpen
  \bibfield  {author} {\bibinfo {author} {\bibfnamefont {J.~J.}\ \bibnamefont
  {Yu}}, \bibinfo {author} {\bibfnamefont {Y.}~\bibnamefont {Yu}}, \bibinfo
  {author} {\bibfnamefont {D.~F.}\ \bibnamefont {Agterberg}}, \ and\ \bibinfo
  {author} {\bibfnamefont {S.}~\bibnamefont {Raghu}},\ }\href@noop {} {\enquote
  {\bibinfo {title} {Theory of the low- and high-field superconducting phases
  of ute$_2$},}\ } (\bibinfo {year} {2023}),\ \Eprint
  {http://arxiv.org/abs/2303.02152} {arXiv:2303.02152 [cond-mat.supr-con]}
  \BibitemShut {NoStop}%
\bibitem [{Sup()}]{Supplemental}%
  \BibitemOpen
  \href@noop {} {}\bibinfo {note} {See Supplemental Materials for the results
  of the $1$+$2$-dimensional periodic Anderson model.}\BibitemShut {Stop}%
\bibitem [{\citenamefont {Yanase}\ \emph {et~al.}(2003)\citenamefont {Yanase},
  \citenamefont {Jujo}, \citenamefont {Nomura}, \citenamefont {Ikeda},
  \citenamefont {Hotta},\ and\ \citenamefont {Yamada}}]{yanase2003review}%
  \BibitemOpen
  \bibfield  {author} {\bibinfo {author} {\bibfnamefont {Y.}~\bibnamefont
  {Yanase}}, \bibinfo {author} {\bibfnamefont {T.}~\bibnamefont {Jujo}},
  \bibinfo {author} {\bibfnamefont {T.}~\bibnamefont {Nomura}}, \bibinfo
  {author} {\bibfnamefont {H.}~\bibnamefont {Ikeda}}, \bibinfo {author}
  {\bibfnamefont {T.}~\bibnamefont {Hotta}}, \ and\ \bibinfo {author}
  {\bibfnamefont {K.}~\bibnamefont {Yamada}},\ }\href {\doibase
  https://doi.org/10.1016/j.physrep.2003.07.002} {\bibfield  {journal}
  {\bibinfo  {journal} {Physics Reports}\ }\textbf {\bibinfo {volume} {387}},\
  \bibinfo {pages} {1} (\bibinfo {year} {2003})}\BibitemShut {NoStop}%
\bibitem [{\citenamefont {Fujibayashi}\ \emph {et~al.}(2023)\citenamefont
  {Fujibayashi}, \citenamefont {Kinjo}, \citenamefont {Nakamine}, \citenamefont
  {Kitagawa}, \citenamefont {Ishida}, \citenamefont {Tokunaga}, \citenamefont
  {Sakai}, \citenamefont {Kambe}, \citenamefont {Nakamura}, \citenamefont
  {Shimizu}, \citenamefont {Homma}, \citenamefont {Li}, \citenamefont {Honda},\
  and\ \citenamefont {Aoki}}]{fujibayashi2023}%
  \BibitemOpen
  \bibfield  {author} {\bibinfo {author} {\bibfnamefont {H.}~\bibnamefont
  {Fujibayashi}}, \bibinfo {author} {\bibfnamefont {K.}~\bibnamefont {Kinjo}},
  \bibinfo {author} {\bibfnamefont {G.}~\bibnamefont {Nakamine}}, \bibinfo
  {author} {\bibfnamefont {S.}~\bibnamefont {Kitagawa}}, \bibinfo {author}
  {\bibfnamefont {K.}~\bibnamefont {Ishida}}, \bibinfo {author} {\bibfnamefont
  {Y.}~\bibnamefont {Tokunaga}}, \bibinfo {author} {\bibfnamefont
  {H.}~\bibnamefont {Sakai}}, \bibinfo {author} {\bibfnamefont
  {S.}~\bibnamefont {Kambe}}, \bibinfo {author} {\bibfnamefont
  {A.}~\bibnamefont {Nakamura}}, \bibinfo {author} {\bibfnamefont
  {Y.}~\bibnamefont {Shimizu}}, \bibinfo {author} {\bibfnamefont
  {Y.}~\bibnamefont {Homma}}, \bibinfo {author} {\bibfnamefont
  {D.}~\bibnamefont {Li}}, \bibinfo {author} {\bibfnamefont {F.}~\bibnamefont
  {Honda}}, \ and\ \bibinfo {author} {\bibfnamefont {D.}~\bibnamefont {Aoki}},\
  }\href {\doibase 10.7566/JPSJ.92.053702} {\bibfield  {journal} {\bibinfo
  {journal} {Journal of the Physical Society of Japan}\ }\textbf {\bibinfo
  {volume} {92}},\ \bibinfo {pages} {053702} (\bibinfo {year} {2023})},\
  \Eprint {http://arxiv.org/abs/https://doi.org/10.7566/JPSJ.92.053702}
  {https://doi.org/10.7566/JPSJ.92.053702} \BibitemShut {NoStop}%
\bibitem [{\citenamefont {Ambika}\ \emph {et~al.}(2022)\citenamefont {Ambika},
  \citenamefont {Ding}, \citenamefont {Rana}, \citenamefont {Frank},
  \citenamefont {Green}, \citenamefont {Ran}, \citenamefont {Butch},\ and\
  \citenamefont {Furukawa}}]{ambika2022}%
  \BibitemOpen
  \bibfield  {author} {\bibinfo {author} {\bibfnamefont {D.~V.}\ \bibnamefont
  {Ambika}}, \bibinfo {author} {\bibfnamefont {Q.-P.}\ \bibnamefont {Ding}},
  \bibinfo {author} {\bibfnamefont {K.}~\bibnamefont {Rana}}, \bibinfo {author}
  {\bibfnamefont {C.~E.}\ \bibnamefont {Frank}}, \bibinfo {author}
  {\bibfnamefont {E.~L.}\ \bibnamefont {Green}}, \bibinfo {author}
  {\bibfnamefont {S.}~\bibnamefont {Ran}}, \bibinfo {author} {\bibfnamefont
  {N.~P.}\ \bibnamefont {Butch}}, \ and\ \bibinfo {author} {\bibfnamefont
  {Y.}~\bibnamefont {Furukawa}},\ }\href {\doibase
  10.1103/PhysRevB.105.L220403} {\bibfield  {journal} {\bibinfo  {journal}
  {Phys. Rev. B}\ }\textbf {\bibinfo {volume} {105}},\ \bibinfo {pages}
  {L220403} (\bibinfo {year} {2022})}\BibitemShut {NoStop}%
\bibitem [{\citenamefont {Ran}\ \emph {et~al.}(2019{\natexlab{b}})\citenamefont
  {Ran}, \citenamefont {Liu}, \citenamefont {Eo}, \citenamefont {Campbell},
  \citenamefont {Neves}, \citenamefont {Fuhrman}, \citenamefont {Saha},
  \citenamefont {Eckberg}, \citenamefont {Kim}, \citenamefont {Graf},
  \citenamefont {Balakirev}, \citenamefont {Singleton}, \citenamefont
  {Paglione},\ and\ \citenamefont {Butch}}]{ran2019extreme}%
  \BibitemOpen
  \bibfield  {author} {\bibinfo {author} {\bibfnamefont {S.}~\bibnamefont
  {Ran}}, \bibinfo {author} {\bibfnamefont {I.-L.}\ \bibnamefont {Liu}},
  \bibinfo {author} {\bibfnamefont {Y.~S.}\ \bibnamefont {Eo}}, \bibinfo
  {author} {\bibfnamefont {D.~J.}\ \bibnamefont {Campbell}}, \bibinfo {author}
  {\bibfnamefont {P.~M.}\ \bibnamefont {Neves}}, \bibinfo {author}
  {\bibfnamefont {W.~T.}\ \bibnamefont {Fuhrman}}, \bibinfo {author}
  {\bibfnamefont {S.~R.}\ \bibnamefont {Saha}}, \bibinfo {author}
  {\bibfnamefont {C.}~\bibnamefont {Eckberg}}, \bibinfo {author} {\bibfnamefont
  {H.}~\bibnamefont {Kim}}, \bibinfo {author} {\bibfnamefont {D.}~\bibnamefont
  {Graf}}, \bibinfo {author} {\bibfnamefont {F.}~\bibnamefont {Balakirev}},
  \bibinfo {author} {\bibfnamefont {J.}~\bibnamefont {Singleton}}, \bibinfo
  {author} {\bibfnamefont {J.}~\bibnamefont {Paglione}}, \ and\ \bibinfo
  {author} {\bibfnamefont {N.~P.}\ \bibnamefont {Butch}},\ }\href
  {https://www.nature.com/articles/s41567-019-0670-X} {\bibfield  {journal}
  {\bibinfo  {journal} {Nature Physics}\ }\textbf {\bibinfo {volume} {15}},\
  \bibinfo {pages} {1250} (\bibinfo {year} {2019}{\natexlab{b}})}\BibitemShut
  {NoStop}%
\bibitem [{\citenamefont {Rosuel}\ \emph {et~al.}(2023)\citenamefont {Rosuel},
  \citenamefont {Marcenat}, \citenamefont {Knebel}, \citenamefont {Klein},
  \citenamefont {Pourret}, \citenamefont {Marquardt}, \citenamefont {Niu},
  \citenamefont {Rousseau}, \citenamefont {Demuer}, \citenamefont {Seyfarth},
  \citenamefont {Lapertot}, \citenamefont {Aoki}, \citenamefont {Braithwaite},
  \citenamefont {Flouquet},\ and\ \citenamefont {Brison}}]{Rosuel2023}%
  \BibitemOpen
  \bibfield  {author} {\bibinfo {author} {\bibfnamefont {A.}~\bibnamefont
  {Rosuel}}, \bibinfo {author} {\bibfnamefont {C.}~\bibnamefont {Marcenat}},
  \bibinfo {author} {\bibfnamefont {G.}~\bibnamefont {Knebel}}, \bibinfo
  {author} {\bibfnamefont {T.}~\bibnamefont {Klein}}, \bibinfo {author}
  {\bibfnamefont {A.}~\bibnamefont {Pourret}}, \bibinfo {author} {\bibfnamefont
  {N.}~\bibnamefont {Marquardt}}, \bibinfo {author} {\bibfnamefont
  {Q.}~\bibnamefont {Niu}}, \bibinfo {author} {\bibfnamefont {S.}~\bibnamefont
  {Rousseau}}, \bibinfo {author} {\bibfnamefont {A.}~\bibnamefont {Demuer}},
  \bibinfo {author} {\bibfnamefont {G.}~\bibnamefont {Seyfarth}}, \bibinfo
  {author} {\bibfnamefont {G.}~\bibnamefont {Lapertot}}, \bibinfo {author}
  {\bibfnamefont {D.}~\bibnamefont {Aoki}}, \bibinfo {author} {\bibfnamefont
  {D.}~\bibnamefont {Braithwaite}}, \bibinfo {author} {\bibfnamefont
  {J.}~\bibnamefont {Flouquet}}, \ and\ \bibinfo {author} {\bibfnamefont
  {J.~P.}\ \bibnamefont {Brison}},\ }\href {\doibase
  10.1103/PhysRevX.13.011022} {\bibfield  {journal} {\bibinfo  {journal} {Phys.
  Rev. X}\ }\textbf {\bibinfo {volume} {13}},\ \bibinfo {pages} {011022}
  (\bibinfo {year} {2023})}\BibitemShut {NoStop}%
\bibitem [{\citenamefont {Kinjo}\ \emph {et~al.}(2023)\citenamefont {Kinjo},
  \citenamefont {Fujibayashi}, \citenamefont {Kitagawa}, \citenamefont
  {Ishida}, \citenamefont {Tokunaga}, \citenamefont {Sakai}, \citenamefont
  {Kambe}, \citenamefont {Nakamura}, \citenamefont {Shimizu}, \citenamefont
  {Homma}, \citenamefont {Li}, \citenamefont {Honda}, \citenamefont {Aoki},
  \citenamefont {Hiraki}, \citenamefont {Kimata},\ and\ \citenamefont
  {Sasaki}}]{Kinjo2023}%
  \BibitemOpen
  \bibfield  {author} {\bibinfo {author} {\bibfnamefont {K.}~\bibnamefont
  {Kinjo}}, \bibinfo {author} {\bibfnamefont {H.}~\bibnamefont {Fujibayashi}},
  \bibinfo {author} {\bibfnamefont {S.}~\bibnamefont {Kitagawa}}, \bibinfo
  {author} {\bibfnamefont {K.}~\bibnamefont {Ishida}}, \bibinfo {author}
  {\bibfnamefont {Y.}~\bibnamefont {Tokunaga}}, \bibinfo {author}
  {\bibfnamefont {H.}~\bibnamefont {Sakai}}, \bibinfo {author} {\bibfnamefont
  {S.}~\bibnamefont {Kambe}}, \bibinfo {author} {\bibfnamefont
  {A.}~\bibnamefont {Nakamura}}, \bibinfo {author} {\bibfnamefont
  {Y.}~\bibnamefont {Shimizu}}, \bibinfo {author} {\bibfnamefont
  {Y.}~\bibnamefont {Homma}}, \bibinfo {author} {\bibfnamefont {D.~X.}\
  \bibnamefont {Li}}, \bibinfo {author} {\bibfnamefont {F.}~\bibnamefont
  {Honda}}, \bibinfo {author} {\bibfnamefont {D.}~\bibnamefont {Aoki}},
  \bibinfo {author} {\bibfnamefont {K.}~\bibnamefont {Hiraki}}, \bibinfo
  {author} {\bibfnamefont {M.}~\bibnamefont {Kimata}}, \ and\ \bibinfo {author}
  {\bibfnamefont {T.}~\bibnamefont {Sasaki}},\ }\href {\doibase
  10.1103/PhysRevB.107.L060502} {\bibfield  {journal} {\bibinfo  {journal}
  {Phys. Rev. B}\ }\textbf {\bibinfo {volume} {107}},\ \bibinfo {pages}
  {L060502} (\bibinfo {year} {2023})}\BibitemShut {NoStop}%
\bibitem [{\citenamefont {Tokiwa}\ \emph {et~al.}(2022)\citenamefont {Tokiwa},
  \citenamefont {Opletal}, \citenamefont {Sakai}, \citenamefont {Kubo},
  \citenamefont {Yamamoto}, \citenamefont {Kambe}, \citenamefont {Kimata},
  \citenamefont {Awaji}, \citenamefont {Sasaki}, \citenamefont {Aoki},
  \citenamefont {Tokunaga},\ and\ \citenamefont
  {Haga}}]{tokiwa2022stabilization}%
  \BibitemOpen
  \bibfield  {author} {\bibinfo {author} {\bibfnamefont {Y.}~\bibnamefont
  {Tokiwa}}, \bibinfo {author} {\bibfnamefont {P.}~\bibnamefont {Opletal}},
  \bibinfo {author} {\bibfnamefont {H.}~\bibnamefont {Sakai}}, \bibinfo
  {author} {\bibfnamefont {K.}~\bibnamefont {Kubo}}, \bibinfo {author}
  {\bibfnamefont {E.}~\bibnamefont {Yamamoto}}, \bibinfo {author}
  {\bibfnamefont {S.}~\bibnamefont {Kambe}}, \bibinfo {author} {\bibfnamefont
  {M.}~\bibnamefont {Kimata}}, \bibinfo {author} {\bibfnamefont
  {S.}~\bibnamefont {Awaji}}, \bibinfo {author} {\bibfnamefont
  {T.}~\bibnamefont {Sasaki}}, \bibinfo {author} {\bibfnamefont
  {D.}~\bibnamefont {Aoki}}, \bibinfo {author} {\bibfnamefont {Y.}~\bibnamefont
  {Tokunaga}}, \ and\ \bibinfo {author} {\bibfnamefont {Y.}~\bibnamefont
  {Haga}},\ }\href@noop {} {\enquote {\bibinfo {title} {Stabilization of
  superconductivity by metamagnetism in an easy-axis magnetic field on
  ute$_2$},}\ } (\bibinfo {year} {2022}),\ \Eprint
  {http://arxiv.org/abs/2210.11769} {arXiv:2210.11769 [cond-mat.supr-con]}
  \BibitemShut {NoStop}%
\bibitem [{\citenamefont {Sakai}\ \emph {et~al.}(2023)\citenamefont {Sakai},
  \citenamefont {Tokiwa}, \citenamefont {Opletal}, \citenamefont {Kimata},
  \citenamefont {Awaji}, \citenamefont {Sasaki}, \citenamefont {Aoki},
  \citenamefont {Kambe}, \citenamefont {Tokunaga},\ and\ \citenamefont
  {Haga}}]{sakai2022field}%
  \BibitemOpen
  \bibfield  {author} {\bibinfo {author} {\bibfnamefont {H.}~\bibnamefont
  {Sakai}}, \bibinfo {author} {\bibfnamefont {Y.}~\bibnamefont {Tokiwa}},
  \bibinfo {author} {\bibfnamefont {P.}~\bibnamefont {Opletal}}, \bibinfo
  {author} {\bibfnamefont {M.}~\bibnamefont {Kimata}}, \bibinfo {author}
  {\bibfnamefont {S.}~\bibnamefont {Awaji}}, \bibinfo {author} {\bibfnamefont
  {T.}~\bibnamefont {Sasaki}}, \bibinfo {author} {\bibfnamefont
  {D.}~\bibnamefont {Aoki}}, \bibinfo {author} {\bibfnamefont {S.}~\bibnamefont
  {Kambe}}, \bibinfo {author} {\bibfnamefont {Y.}~\bibnamefont {Tokunaga}}, \
  and\ \bibinfo {author} {\bibfnamefont {Y.}~\bibnamefont {Haga}},\ }\href
  {\doibase 10.1103/PhysRevLett.130.196002} {\bibfield  {journal} {\bibinfo
  {journal} {Phys. Rev. Lett.}\ }\textbf {\bibinfo {volume} {130}},\ \bibinfo
  {pages} {196002} (\bibinfo {year} {2023})}\BibitemShut {NoStop}%
\bibitem [{Sue()}]{Suetsugu2023}%
  \BibitemOpen
  \href@noop {} {}\bibinfo {note} {S. Suetsugu {\it et al.},
  submitted.}\BibitemShut {Stop}%
\bibitem [{\citenamefont {Matsumura}\ \emph {et~al.}(2023)\citenamefont
  {Matsumura}, \citenamefont {Fujibayashi}, \citenamefont {Kinjo},
  \citenamefont {Kitagawa}, \citenamefont {Ishida}, \citenamefont {Tokunaga},
  \citenamefont {Sakai}, \citenamefont {Kambe}, \citenamefont {Nakamura},
  \citenamefont {Shimizu}, \citenamefont {Homma}, \citenamefont {Li},
  \citenamefont {Honda},\ and\ \citenamefont {Aoki}}]{matsumura2023large}%
  \BibitemOpen
  \bibfield  {author} {\bibinfo {author} {\bibfnamefont {H.}~\bibnamefont
  {Matsumura}}, \bibinfo {author} {\bibfnamefont {H.}~\bibnamefont
  {Fujibayashi}}, \bibinfo {author} {\bibfnamefont {K.}~\bibnamefont {Kinjo}},
  \bibinfo {author} {\bibfnamefont {S.}~\bibnamefont {Kitagawa}}, \bibinfo
  {author} {\bibfnamefont {K.}~\bibnamefont {Ishida}}, \bibinfo {author}
  {\bibfnamefont {Y.}~\bibnamefont {Tokunaga}}, \bibinfo {author}
  {\bibfnamefont {H.}~\bibnamefont {Sakai}}, \bibinfo {author} {\bibfnamefont
  {S.}~\bibnamefont {Kambe}}, \bibinfo {author} {\bibfnamefont
  {A.}~\bibnamefont {Nakamura}}, \bibinfo {author} {\bibfnamefont
  {Y.}~\bibnamefont {Shimizu}}, \bibinfo {author} {\bibfnamefont
  {Y.}~\bibnamefont {Homma}}, \bibinfo {author} {\bibfnamefont
  {D.}~\bibnamefont {Li}}, \bibinfo {author} {\bibfnamefont {F.}~\bibnamefont
  {Honda}}, \ and\ \bibinfo {author} {\bibfnamefont {D.}~\bibnamefont {Aoki}},\
  }\href@noop {} {\enquote {\bibinfo {title} {Large reduction in the $a$-axis
  knight shift on ute$_2$ with $t_{\rm c}$ = 2.1 k},}\ } (\bibinfo {year}
  {2023}),\ \Eprint {http://arxiv.org/abs/2305.01200} {arXiv:2305.01200
  [cond-mat.supr-con]} \BibitemShut {NoStop}%
\bibitem [{\citenamefont {Leggett}(1975)}]{Leggett1975}%
  \BibitemOpen
  \bibfield  {author} {\bibinfo {author} {\bibfnamefont {A.~J.}\ \bibnamefont
  {Leggett}},\ }\href {\doibase 10.1103/RevModPhys.47.331} {\bibfield
  {journal} {\bibinfo  {journal} {Rev. Mod. Phys.}\ }\textbf {\bibinfo {volume}
  {47}},\ \bibinfo {pages} {331} (\bibinfo {year} {1975})}\BibitemShut
  {NoStop}%
\bibitem [{\citenamefont {Tei}\ \emph {et~al.}(2023)\citenamefont {Tei},
  \citenamefont {Mizushima},\ and\ \citenamefont {Fujimoto}}]{Tei2023}%
  \BibitemOpen
  \bibfield  {author} {\bibinfo {author} {\bibfnamefont {J.}~\bibnamefont
  {Tei}}, \bibinfo {author} {\bibfnamefont {T.}~\bibnamefont {Mizushima}}, \
  and\ \bibinfo {author} {\bibfnamefont {S.}~\bibnamefont {Fujimoto}},\ }\href
  {\doibase 10.1103/PhysRevB.107.144517} {\bibfield  {journal} {\bibinfo
  {journal} {Phys. Rev. B}\ }\textbf {\bibinfo {volume} {107}},\ \bibinfo
  {pages} {144517} (\bibinfo {year} {2023})}\BibitemShut {NoStop}%
\bibitem [{\citenamefont {Kawamura}(2019)}]{kawamura2019fermisurfer}%
  \BibitemOpen
  \bibfield  {author} {\bibinfo {author} {\bibfnamefont {M.}~\bibnamefont
  {Kawamura}},\ }\href@noop {} {\bibfield  {journal} {\bibinfo  {journal}
  {Computer Physics Communications}\ }\textbf {\bibinfo {volume} {239}},\
  \bibinfo {pages} {197} (\bibinfo {year} {2019})}\BibitemShut {NoStop}%
\end{thebibliography}%

\clearpage

\begin{center}
 {\large \textmd{Supplemental Materials:} \\[0.3em]
 {\bfseries Magnetism and superconductivity in mixed-dimensional periodic Anderson model for UTe$_{2}$}}
\end{center}

\begin{center}
\section{$1$+$2$-dimensional periodic Anderson model}
\end{center}

We show the results of the $1$+$2$-dimensional periodic Anderson model to compare the model with the $1$+$3$-dimensional periodic Anderson model discussed in the main text.
Combining the results of $1$+$2$- and $1$+$3$-dimensional models, we reveal general properties of the mixed-dimensional periodic Anderson model.
In the $1$+$2$-dimensional periodic Anderson model, the hopping integral of $f$ electrons along of the $k_z$ direction is taken as zero, $t_{fz}=0$.
The other parameters are the same as the $1$+$3$-dimensional model.
We study magnetic fluctuation and superconducting instability by changing the $f$-electron level $\Delta_f$ from $0.05$ to $0.13$, as we have done for the $1$+$3$-dimensional model in the main text. 

\begin{figure}[htbp]
    \centering
\includegraphics[width=0.45\textwidth]{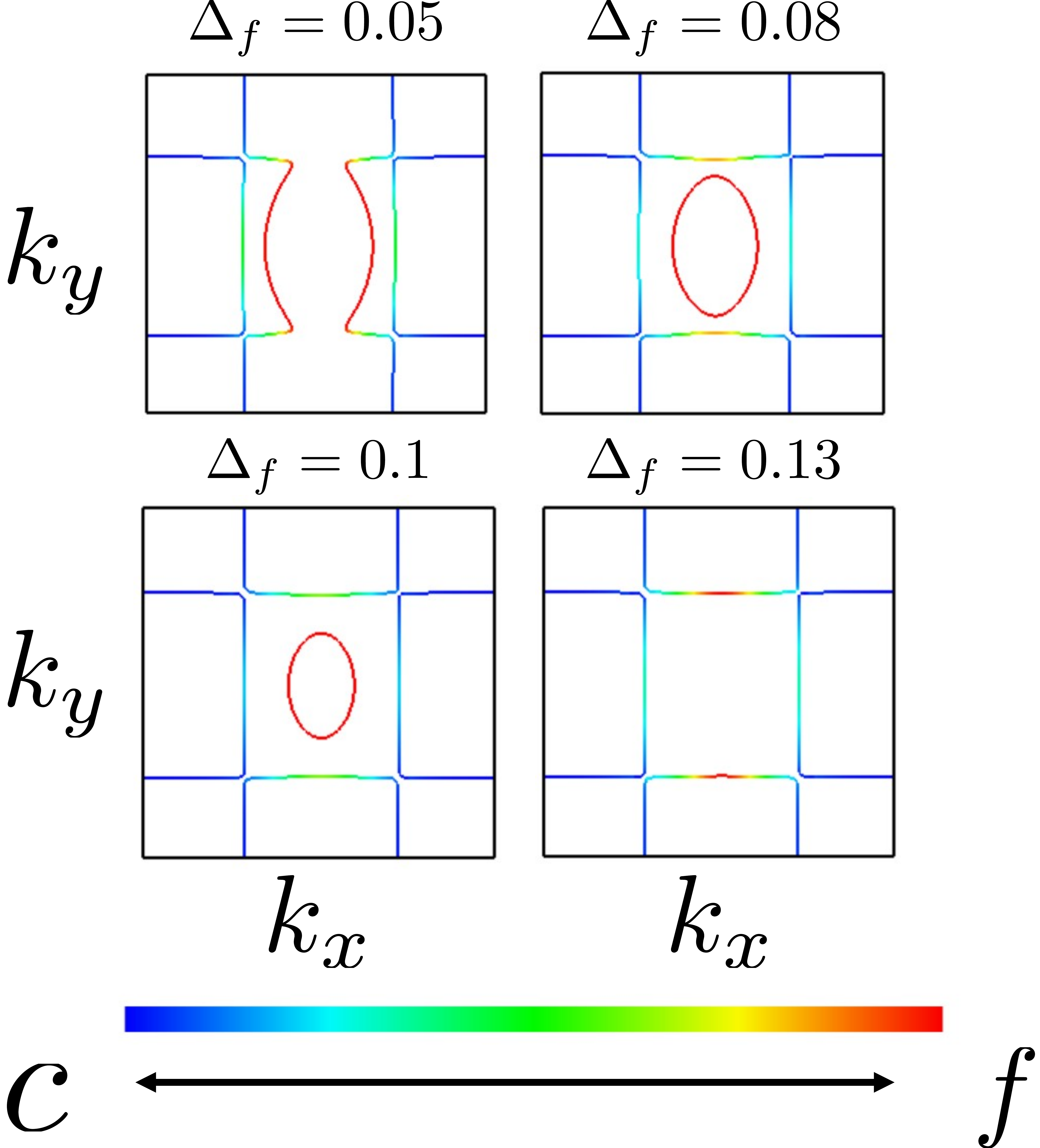}
    \caption{
    Fermi surfaces of the $1$+$2$-dimensional periodic Anderson model. The color indicates the weight of $f$ electrons. 
    }
\label{fig:2D_Fermi}
\end{figure}

First, Fermi surfaces of the $1$+$2$-dimensional periodic Anderson model are shown in Fig.~\ref{fig:2D_Fermi}.
They are equivalent to the Fermi surfaces of the $1$+$3$-dimensional model at $k_{z}=\pi$.

\begin{figure}[htbp]
    \centering
\includegraphics[width=0.45\textwidth]{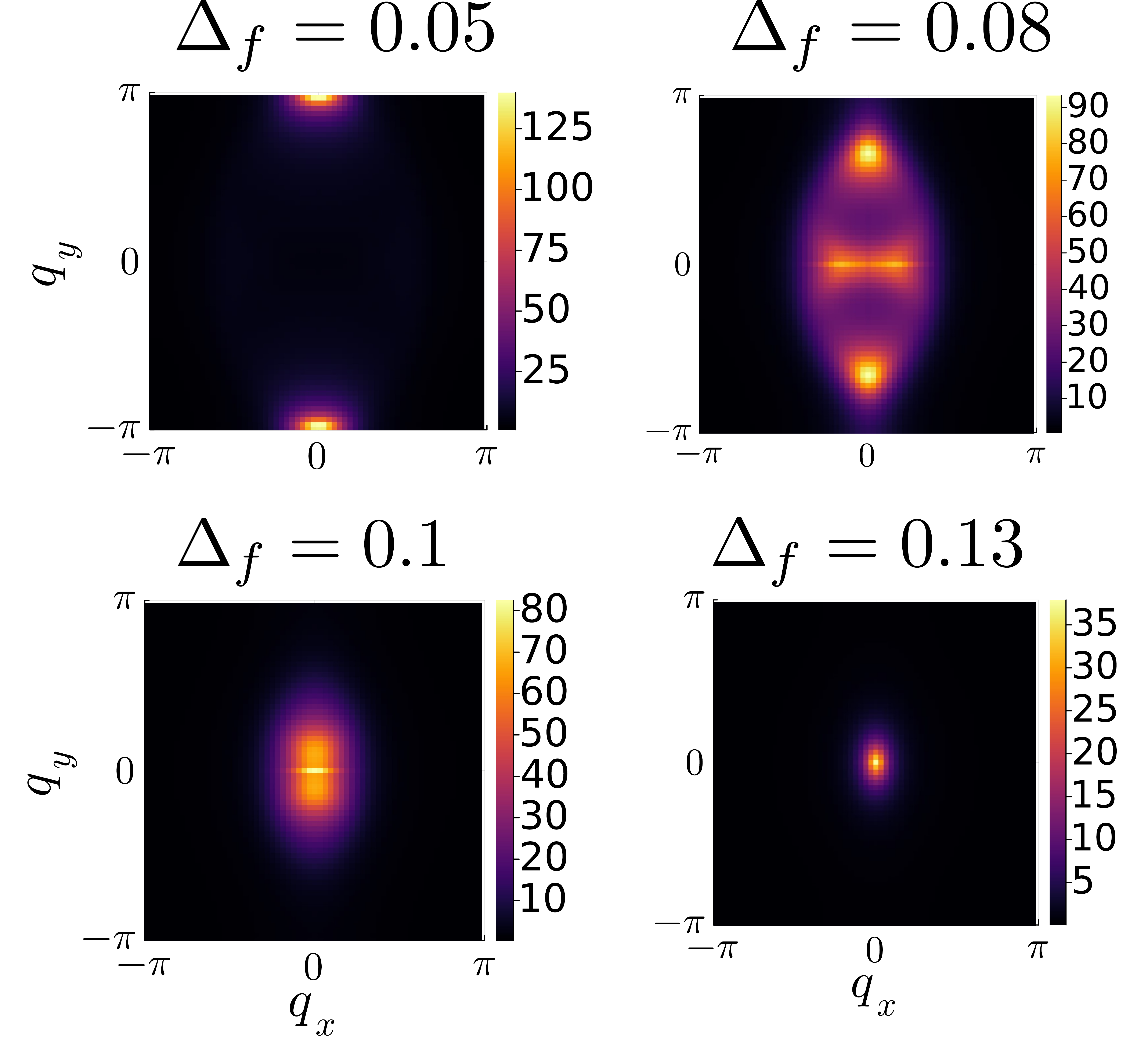}
    \caption{
    Static spin susceptibility $\chi_s(\bm q) \equiv \chi_s(\bm q,i\Omega_m=0)$ for various $f$-electron levels $\Delta_f$.
    }
\label{fig:2D_susc}
\end{figure}

Second, we show the spin susceptibility in Fig.~\ref{fig:2D_susc}.
We see antiferromagnetic fluctuation with the wave vector $\bm{q}=(0,\pi)$ when $\Delta_f=0.05$, the same as in the $1$+$3$-dimensional model.
In the $1$+$2$-dimensional model, ferromagnetic fluctuation is not obtained for $\Delta_f$ ranging from $0.05$ to $0.07$, different from the $1$+$3$-dimensional model where ferromagnetic fluctuation appears above $\Delta_f =0.05$.
The wave vector of antiferromagnetic fluctuation changes with increasing $\Delta_f$ and crossovers to ferromagnetic fluctuation around $\Delta_f =0.08$.
Thus, the region where antiferromagnetic fluctuation dominates is broader than that in the $1$+$3$-dimensional model.
This is because the two-dimensional model has better nesting properties than the three-dimensional model and the cylindrical Fermi surface is less likely to show ferromagnetic fluctuation than the spherical Fermi surface.

\begin{table}[hbtp]
  \caption{Irreducible representations and basis functions in the $1$+$2$-dimensional model.}
\label{table:representation}
\vspace{3mm}
  \centering
  \begin{tabular}{|c|c|c|c|c|}
    \hline
    IR($D_{2h}$) & $A_g$ & $B_{1g}$ & $B_{2u}$ & $B_{3u}$ \\
    \hline
    Basis function & $1$ & $k_x k_y$ & $k_y$ & $k_x$\\
    \hline
  \end{tabular}
\end{table}

\begin{figure}[htbp]
    \centering
\includegraphics[width=0.45\textwidth]{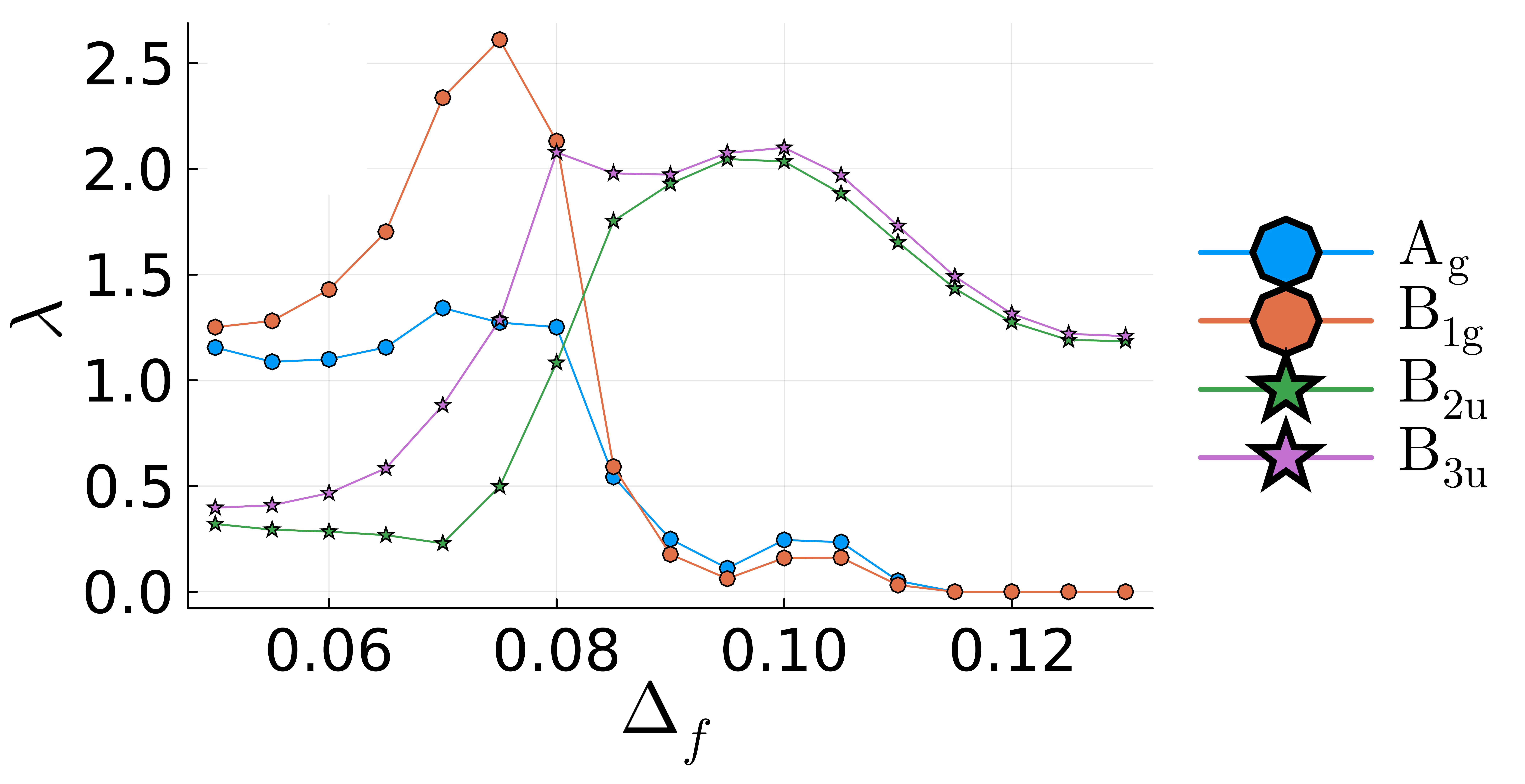}
    \caption{
    Eigenvalues of \'{E}liashberg equation for superconductivity in the $1$+$2$-dimensional periodic Anderson model. $\Delta_{f}$ dependence is shown for each irreducible representation of the $D_{2h}$ point group (Table~\ref{table:representation}). 
    }
\label{fig:2D_lambda}
\end{figure}

\begin{figure}[htbp]
    \centering
\includegraphics[width=0.45\textwidth]{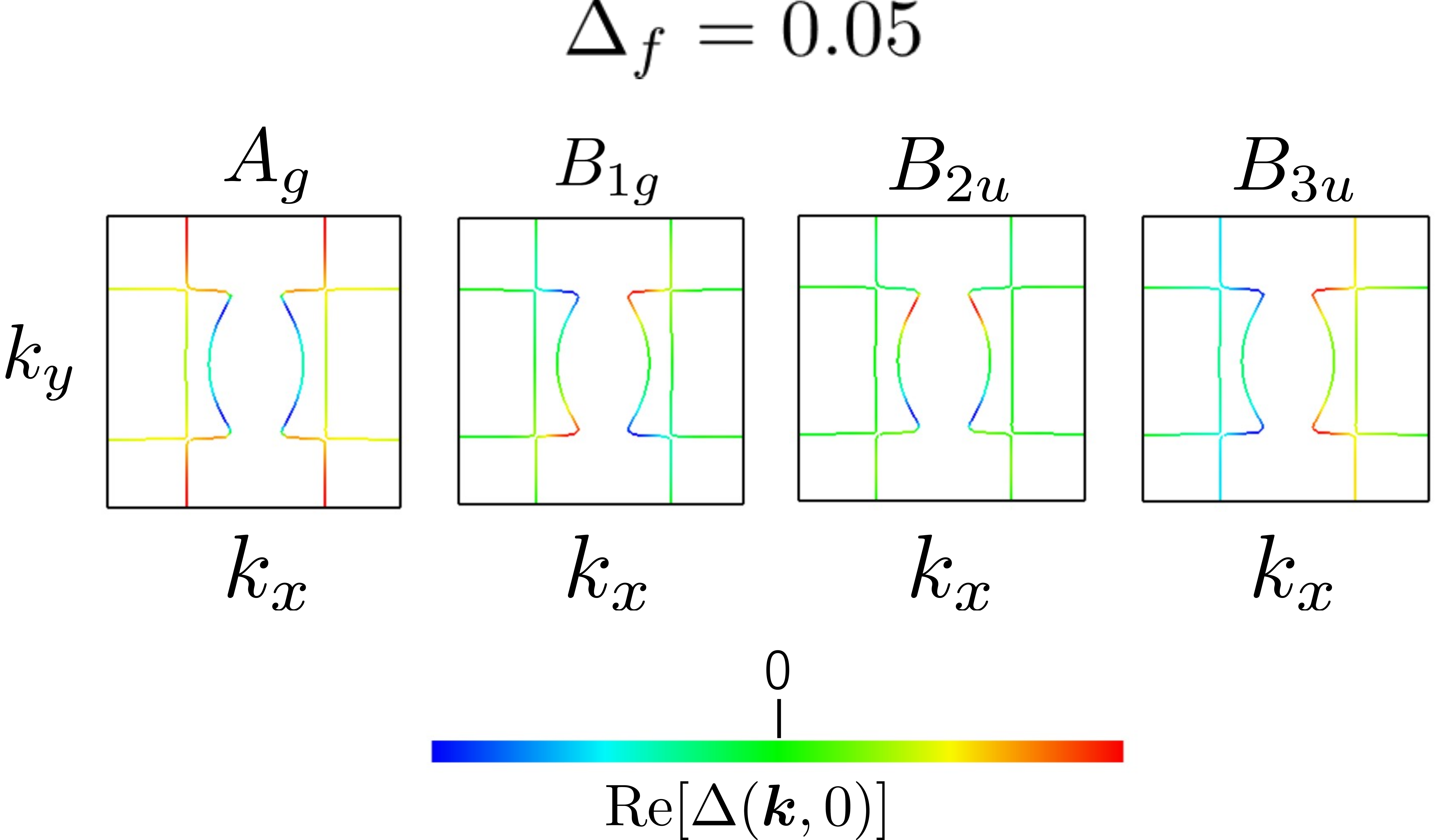}
    \caption{
    Superconducting gap function in the $1$+$2$-dimensional periodic Anderson model for $\Delta_f=0.05$.
    }
\label{fig:2D_gap}
\end{figure}

Next, we discuss superconductivity.
The gap function in the $1$+$2$-dimensional model does not have $k_z$ dependence, and therefore, the superconducting states are classified by the irreducible representations in Table~\ref{table:representation}. 
The eigenvalues of \'{E}liashberg equation as functions of $\Delta_f$ are shown in Fig.~\ref{fig:2D_lambda}.
We see the stable superconducting state changes from the $B_{1g}$ representation to the $B_{2u}$ and $B_{3u}$ representations at around $\Delta_f=0.08$, associated with antiferromagnetic to ferromagnetic crossover in magnetic fluctuation.
For $\Delta_f <0.08$, the spin-singlet superconducting state of $B_{1g}$ representation is stabilized by antiferromagnetic fluctuation.
On the other hand, for $\Delta_f >0.08$, spin-triplet superconducting states of $B_{2u}$ and $B_{3u}$ representations are stabilized by ferromagnetic fluctuation, and spin-singlet $B_{1g}$ and $A_{g}$ states are almost suppressed.
It is stressed that antiferromagnetic fluctuation and spin-triplet $p$-wave superconductivity do not coexist in the $1$+$2$-dimensional model, in contrast to the $1$+$3$-dimensional model.
The gap function on the Fermi surface for $\Delta_f=0.05$ is shown in Fig.~\ref{fig:2D_gap}.

\end{document}